\documentclass{amsproc}

\newtheorem{theorem}{Theorem}[section]
\newtheorem{lemma}[theorem]{Lemma}

\theoremstyle{definition}
\newtheorem{definition}[theorem]{Definition}

\theoremstyle{remark}

\numberwithin{equation}{section}
\usepackage{graphicx}
\usepackage{verbatim}
\begin{document}

\title{Genetic-metabolic networks can be modeled as toric varieties.}

%    Remove any unused author tags.

%    author one information
\author{Marco Polo Castillo-Villalba}
\address{Program of Computational Genomics, Center for Genomic Sciences UNAM, C.P. 62210, Cuernavaca, Morelos, Mexico.}
%\curraddr{}
\email{mpolovillalba@gmail.com}
%\thanks{}

%    author two information
\author{Laura G\'{o}mez-Romero}
%\address{Program of Computational Genomics, Center for Genomic Sciences UNAM, C.P. 62210, Cuernavaca, Morelos, Mexico.}
%\curraddr{}
\email{lawrah20@gmail.com}
%\thanks{}
\author{Julio Collado-Vides}
%\address{Program of Computational Genomics, Center for Genomic Sciences UNAM, C.P. 62210, Cuernavaca, Morelos, Mexico.}
%\curraddr{}
\email{colladojulio@gmail.com}

\subjclass[2010]{Primary }


\date{7/8/2017}

%\dedicatory{}

\begin{abstract}
 The mathematical modelling of genetic-metabolic networks is of out most importance in the field of systems biology. Different formalisms and a huge variety of classical mathematical tools have been used to describe and analyse such networks. Michael A. Savageau defined a formalism to model genetic-metabolic networks called \textit{S-Systems}, \cite{MASavageau1976}, \cite{MASavageauVoit1987}, \cite{MASavageau2001}. There is a limit in the number of nodes that can be analysed when these systems are solved using classical numerical methods such as non-linear dynamic analysis and linear optimization algorithms. We propose to use toric algebraic geometry to solve S-systems. In this work we prove that \textbf{S-systems are toric varieties} and that as a consequence \textit{Hilbert basis} can be used to solve them. This is achieved by applying two theorems, proved here, the theorem about Embedding of S-Systems in toric varieties and the theorem about Toric Resolution on S-Systems. In addition, we define the realization of \textit{minimal phenotypic polytopes}, \textit{phenotypic toric ideals} and \textit{phenotypic toric varieties} as a generalization of the \textit{phenotypic polytopes} described by M. Savageau, \cite{LomnitzSavageau2013}, \cite{LomnitzSavageau2014}, \cite{LomnitzSavageau2015}. Also we studied a synthetic oscillatory network of two genes under diverse environmental control modes, we reproduce S. Savageau results and additionally we obtain an invariant sustained oscillatory region.
 %Also we reproduce a particular case of phenotypic polytope studied and reported by M. Savageau \cite{LomnitzSavageau2014}, which is the case of a design of synthetic oscillating network of two genes, submitted under diverse conditions of environmental control modes, obtaining a sustained oscillating region, this result is achieved under the focus of phenotypic toric varieties. 
  The implications of the results here presented is that, in principle, they will facilitate solving large scale genetic-metabolic networks by means of toric algebraic geometry tools as well as facilitate elucidating their dynamics.

 %In this work we give a brief introduction to the fundamental concepts in combinatorial toric algebraic geometry and posteriori, we apply these tools to \textit{gene regulatory-metabolic networks} in the formalism of \textit{S-systems} studied by M. Savageau \cite{MASavageauVoit1987}. The principal contribution in our work is the application of the Theorems \ref{Theorem 1.} and \ref{Theorem 2.} proved here. They permit us to make a formalization of Biochemical Systems particularly on gene networks, \cite{MASavageau1976}, in the theory of \textit{phenotypic polytopes} and \textit{S-systems}, defined and developed by M. Savageau and J. Lomintz, refs \cite{MASavageauVoit1987}, \cite{MASavageau2001}, \cite{LomnitzSavageau2013}, \cite{LomnitzSavageau2014}, \cite{LomnitzSavageau2015}. We show that \textbf{S-systems can be described as toric varieties}. We define the realization of minimal phenotypic polytopes
%using \textit{Hilbert basis}, this computational technique we will permit us to model gene networks of large-scale and to solve its dynamics; also we define a \textit{phenotypic variety} and \textit{phenotypic toric ideals}; in our next work, we will reproduce the conserved phenotypic information of a network with an oscillatory circuit design previously studied by M. Savageau.

\end{abstract}
\maketitle
\section{Introduction.}
The modelling of genetic-metabolic networks allows us to simulate and study the mechanisms of gene regulation, and its interplay with the environment. Gene regulation can be represented as an interaction graph in which each molecule is represented as a node and the interaction between them as links on the graph. The molecules depicted can be genes,  mRNA's, proteins or metabolites. Every interaction can represent a catalytic event, the repression or activation of a transcription factor, or the formation of a complex.

The lack of information about the kinetic parameters and the orders of the biochemical reactions involved in gene regulation and metabolic processes imposes constraints to solve analytically networks with a huge amount of nodes. Nevertheless, there are mathematical tools to approximate the parameter space. For instance, the parameter-space for an S-system can be delimited by linear optimization numerical techniques by identifying parameters intervals in order to approximate the parameter-space.

In the present study, we propose that a genetic-metabolic network, represented as a set of S-systems, can be solved using tools from algebraic geometry. First, we enunciate two theorems to formalize the S-Systems. In these theorems we show that the S-Systems are phenotypic toric varieties. We prove that a phenotypic toric variety is a generalization of a phenotypic polytope. Also, we define a lattice polytope, which is another combinatorial object associated with an S-system and which is required to do the geometrical realization of these algebraic objects. Also, we reproduce the solutions obtained for a set of S-Systems associated with a genetic-metabolic network in steady-state, previously studied by M. Savageau.

The first part of the work contains some background needed for the proof of our theorems. This includes a few concepts from Convex Geometry, such as Gordans Lemma, and also some Combinatorial Toric Algebraic Geometry. In particular, we state the definition of Toric Variety as an Algebraic Afine Scheme, since this definition is important in our formalization of S-Systems. Also, we introduce the main concepts of power law modelling of  biochemical systems (power law, dominant terms, phenotypic polytope), specifically in genetic-metabolic networks as studied by M. Savageau, \cite{MASavageau1976}. We prove that S-systems are toric varieties , both, assuming steady-state by demonstrating the Theorem 4.4-(embedding of S-Systems in toric varieties ), and beyond steady state by demonstrating Theorem 4.5-(toric resolution on S-Systems ).

In the second part, we study a gene network corresponding to a synthetic oscillator studied by M. Savageau \cite{LomnitzSavageau2014}. The network is composed of two genes whose expression is regulated by two transcription factors (TF's). The system is studied under different control modes, i.e; two negative feedback loops, two positive feedback loops, and a nested positive and negative feedback loops. The solution for this system shows that certain geometrical regions of the parameter-space are associated to each gene regulation control mode. Interestingly, most of the geometrical locus of the parameter-space is preserved, and this persistent region is associated with the oscillatory solution of the model. This persistent region is known as a phenotypic polytope, and the oscillatory solution is also called an oscillatory phenotype. Analytically, an oscillatory solution is a complex linear combination of sin and cosine functions varying over time, whose values represent the change of concentrations over time for every biochemical species in the network. 

There are several important implications from  the formalization of the S-Systems as toric varieties. An embedding of any S-System into an algebraic torus must exist. A phenoptypic toric variety, which is an object more general than a phenotypic polytope, can be constructed from minimal objects called phenotypic toric ideals. The phenotypic toric ideals can be computed using Hilbert basis. The toric generators contain all the solution space associated with the S-system; because of this, the oscilattory phenotypes seen by Savageau are reproduced  and some new phenotypic polytopes are found. Finally, the toric generators and the use of Hironaka-Atiyah's theorem allow the study of the dynamics of any set of S-Systems.

\section{Combinatorial Toric Algebraic Geometry Background.}

In this section, we explain the background from combinatorial toric algebraic geometry, which we will need for the proof of our main theorems in section 3. Set $ M \, \subset \, \mathbb{R}^{n} $, by conv(\textbf{M}), we mean the convex hull of $ M $ , which is the set of all convex combinations of elements of $ M $. Given a finite subset M of $\mathbb{R}^{n}$, a \textbf{convex polytope} or \textbf{polytope} is the convex hull of M.
\begin{definition}\label{Definition 1.0} \cite{DCoxLittleSchenck2010}. A \textbf{lattice} $ N $ is a free abelian group of finite rank, and if its rank is $ n \, \in \, \mathbb{N} $, then $ N $ is isomorphic to $\mathbb{Z}^{n} $.\end{definition}
%Given $ M \, \subset \, \mathbb{R}^{n} $, by conv\textbf{M} we mean the hull convex of $ M $ , which is the set of all convex combinations of elements of $ M $. Moreover, if $ M $ is a finite set then conv\textbf{M} is called a \textbf{convex polytope} or \textbf{polytope}.\\

In particular for the case of each biochemical network, is determined by kinetic orders of reaction, which are determined by dissociation constants that are related to the concentrations of reactants to the substrates and products. We define a \textbf{condition lattice vector} $ p_{i}=(g_{i1},...,g_{im},h_{i1},...,h_{im}) \, \in\,\mathbb{Z}^{2m} $ for some integer $ i $; in the case that the $ g_{ij}$ and $ h_{ij} $ are rational numbers, we always can multiply them by a real number $ \lambda \, \in \, \mathbb{R} $, such that $ \lambda*p_{i}=\lambda*(g_{i1},...,g_{im},h_{i1},...,h_{im})  \, \in\,\mathbb{Z}^{2m} $, where each term is related to enviromental conditions related to gene regulation. \\
The set of all condition lattice vectors $ p_{i} \, \in\,\mathbb{Z}^{2m} $ form a finite convex hull set, hence a convex polytope. We are interested in the algebraic geometry properties of these objects.\\

\begin{definition}\label{Definition 1.1} \cite{GEwald1993}. A semi-group, is a non empty set \textbf{S} with an associative operation, 
\begin{center}
$ + :  $\textbf{S}$ \times $\textbf{S}$ \longrightarrow $\textbf{S}.
\end{center}
A semi-group is named a \textbf{monoid}, its operation is commutative and if there exists a neutral element, i.e. an element $ 0 \, \in \,  $ \textbf{S}, where, $ s+0 =s,\, \forall \, s \, \in \, $ \textbf{S}, and this satisfies a cancellation law.\end{definition}
\begin{definition}\label{Definition 1.2.}
Given a cone $ \sigma=Con(p_{1},...,p_{n}) \, \subseteq \, \mathbb{R}^{n}$, if it is generated by lattice vectors, i.e., each $ p_{i} \,\in  \, \mathbb{Z}^{n}$ for all $ i $, we say that it is a \textbf{lattice cone}. And the convex hull of these lattice vectors, $ P= $conv$(p_{1},...,p_{n})$, we call a \textbf{lattice-polytope}. 

%is a hull convex, whose elements form a free abelian group. If the group is of rank $ n $, then it is isomorphic to $ \mathbb{Z}^{n} $; given the case for $ p_{i} \, \in \, \mathbb{Z}^{n} $ with $ i=1,...,n $, we also write $ \sigma=Con(p_{1},...,p_{n}) $ and we say that it is the lattice cone generated by the lattice vectors $ p_{i} $; see other definitions in: \cite{DCoxLittleSchenck2010} and \cite{GEwald1993}.
\end{definition}

\textbf{Lemma 2.1. \cite{GEwald1993}.} Let $ \sigma\, \subset \, \mathbb{R}^{n} $ be a lattice cone, then $ \sigma \cap \mathbb{Z}^{n} $ is a monoid, see \cite{GEwald1993}.\\

\textbf{Lemma 2.2.} \textbf{(Gordan's Lemma)}. \cite{GEwald1993}. Let $ \sigma \, \subset \, \mathbb{R}^{n} $ be a lattice cone, then the monoid $ \sigma \cap \mathbb{Z}^{n} $ is finitely generated.\\
Now given the set of condition lattice vectors, $ p_{i}=(g_{i1},...,g_{im},h_{i1},...,h_{im}) \, \in\,\mathbb{Z}^{2m} $, with $ i=1,...,n $, we construct the lattice cone $ \sigma_{i} $. For some $ i $ and its associated monoid $S_{\sigma} = \sigma^{\vee} \cap \mathbb{Z}^{n}$ only intersecting  with $\mathbb{Z}^{n}  $ for lemma 2.1 (Gordan's lemma)
this monoid is therefore finitely generated, in the perspective of biochemical reactions there exists a set of minimal reactions associated to this minimal set (Hilbert basis) of kinetic orders, these minimal sets give us an interesting way to solve large-scale metabolic networks.\\

For topological reasons of completeness and compactness we will work not in the cone but in the dual cone associated to it. The definition of the dual cone is as follows.

%Our interest is to define the dual lattice cone, since the geometrical realizations in our work are made in the dual lattice cone, by topological reasons of completeness and compactness in the toric variety, so we define the dual lattice cone as follows:

\begin{definition}\label{Definition 1.3.} The dual lattice cone $\sigma^{\vee} \,\subseteq \, \mathbb{Z}^{m} $ associated to the cone $ \sigma \, \subseteq \, \mathbb{Z}^{n}$, is given by,
\begin{center}
$\sigma^{\vee} = \lbrace m \in \sigma: \langle m,u \rangle > 0, \, \forall u \, \in \sigma \rbrace $.
\end{center}
And also the equality of sets is true; $(\sigma^{\vee})^{\vee}= \sigma $.
\end{definition}
Also we can see of the definition, that if $ \sigma \,\subseteq \, \mathbb{Z}^{m} $ is a lattice cone, so $ \sigma^{\vee}  $ it is, and the same way the previous lemmas are applied on it. \\
Now we introduce one the most important concept in this work, as we see in the following.\\

%By means of these lemmas, we enunciate the following results.
%Definition 2.3. An affine toric variety is an irreducible affine variety $ X $
%containing a torus $ T_{N} \simeq (\mathbb{C}^{*})^{n} $ as Zariski open subset, such that the action of $ T_{N} $ on itself, is extended to an algebraic action of; that is, there exists a morphism from $ T_{N}\otimes X $ to $ X $, see ref. \cite{DCoxLittleSchenck2010}.
%Let $ \sigma $ be a lattice cone, the affine algebraic scheme: $ X_{\sigma} =$\textbf{Spec}$ (R_{\sigma}) $.
%is called abstract toric affine variety or embedding of torus.
The spectrum \textbf{Spec} of a ring $ R $, or algebraic affine scheme \textbf{Spec}$ (R) $ consists of all prime ideals of $ R $. In this work, we focus on the ring of polynomials of Laurent; i.e., $ \mathbb{C}[x,x^{-1}] $. And so, we construct the coordinate ring $ R_{\sigma} $
in the following manner. We take the support of any polynomial $ g \,\in \, \mathbb{C}[x,x^{-1}] $, and denote $ supp(g) $, which forms a lattice cone, called the Newton polyhedron or exponent-space. For some cone lattice $ \sigma \, \subset \, \mathbb{R}^{n} $, we have $R_{\sigma}= \left\lbrace g \, \in \, \mathbb{C}[x,x^{-1}] \, : \, supp(g) \, \subset \, \sigma^{\vee} \right\rbrace   $.

\begin{definition}\label{Definition2.} Let $\sigma$  be a lattice cone, the   \textbf{affine algebraic scheme:}
\begin{center}
$  X_{\sigma}=$ \textbf{Spec} $ (R_{\sigma}) $.
\end{center}
is called an \textbf{abstract toric affine variety} or \textbf{embedding of torus.}\\
\end{definition}

\section{Toric resolution and toric morphisms.}

Hibert basis in an important tool in Algebraic Geometry, \cite{BSturmfels1996}. In our case, 
the Hilbert basis associated to a monoid $ N\, \subset \, \mathbb{Z}^{n} $, will enable us to compute explicitly a toric resolution in our main theorem. Also, its utilization will have important practical applications in the modelling of the dynamic solution and the steady-state of genetic-metabolic networks.

%the Hilbert basis associated to a monoid will enable us to compute explicitly a toric resolution in our main theorem (the definition of toric resolution will be given in this section).
%\cite{BSturmfels1996}, \cite{DCoxLittleSchenck2010}. The utilization of Hilbert basis associated to a monoid $ N\, \subset \, \mathbb{Z}^{n} $ will be important for the practical applications in the dynamical solution and the steady-state in the modelling of gene metabolic-networks, see \cite{BSturmfels1996}.\\

%Let $N\, \simeq \, \mathbb{Z}^{n}$ be a lattice and set $ M=N^{\vee} $ its dual lattice which consist in the group of all the homomorphism of $ N $ to $ M $, is possible to see that the a dual lattice is yield from Monoid $ \sigma^{\vee} $. Let $ \sigma $ be a lattice cone defined in $ N $ and let $ \sigma^{\vee} $ be its dual cone in $ M $ . 
%%%Now given the set of condition lattice vectors, $ p_{i}=(g_{i1},...,g_{im},h_{i1},...,h_{im}) \, \in\,\mathbb{Z}^{2m} $, with $ i=1,...,n $, we construct the lattice cone $ \sigma_{i} $. For some $ i $ and its associated monoid $S_{\sigma} = \sigma^{\vee} \cap \mathbb{Z}^{n}$ only intersecting  with $\mathbb{Z}^{n}  $ for lemma 2.1 (Gordan's lemma)
%this monoid is therefore finitely generated, in the perspective of biochemical reactions there exists a set of minimal reactions associated to this minimal set (Hilbert basis) of kinetic orders, these minimal sets give us an interesting way to solve large-scale metabolic networks.\\

\begin{lemma}\label{Lemma 3.0.}\textbf{(Hilbert basis)}, \cite{BSturmfels1996}, \cite{DCoxLittleSchenck2010}, \cite{DCoxLittleOshea2004}.  Set $ \sigma \, \subseteq \, N $, then $ \sigma $ is an n-dimensional cone if and only if it is a strongly convex cone; i.e., $ \sigma \cap (-\sigma^{\vee}) ={0}$. In this case the monoid $ S_{\sigma} $ has a finite minimal set of generators $ H\, \subseteq \, M \, \simeq \, \mathbb{Z}^{d} $. \end{lemma}

In toric algebraic geometry it is a common practice to compute toric ideals from Hilbert basis. We will use the singular program for these computations. The following proposition and definition, will allow us to identify when a set of binomials are toric. These kind of binomials are important in our work, because the genetic metabolic networks can be represented in this way. 
%And we only give the proposition 3.0\label{Proposition 2.0.} and definition \label{Definition 2.1.3.}3.1 to verify when a binomial is toric or generate a toric ideal. \\

%Is a standard result of toric algebraic geometry to compute toric ideals from Hilbert basis, and we use for these computations the singular program specialized in computational algebraic geometry, [], and only we give the \label{Proposition 2.0.} and definition \label{Definition 2.1.3.} to verify when a binomial is toric or generate a toric ideal; since that our work has as object to solve the dynamic of genetic metabolic-networks by means these kind of binomials, also they will be used in the proofs of the last theorems.\\

\textbf{Proposition 3.0.}\label{Proposition 2.0.} \cite{DCoxLittleSchenck2010}. Let  $ I $ be an ideal of the affine toric variety $ X_{\sigma} \, \subseteq \, \mathbb{C}^{n}$. Then this ideal has a representation given by:
\begin{center}
$  I(X_{\sigma})= \langle \, t^{l_{+}} - t^{l_{-}} \vert l \, \in L \, \rangle =\langle \, t^{\alpha} - t^{\beta} \vert \alpha - \beta \, \in L, \, \alpha, \, \beta \in \, \mathbb{Z}_{+}^{n} \, \rangle $,
\end{center}
where $ L $ is the kernel of the following morphism $ a_{i} \longrightarrow t^{a_{i}}$ with $ a_{i} \, \in \, \mathbb{Z}^{m}  $, $i=1,...,n  $. 

%where $ L $ is the kernel of the following morphism $ a_{i} \longrightarrow t^{a_{i}} \longrightarrow \mathbb{Z}^{n} \longrightarrow M $ and $ M $ is a monoid  such that $ M \simeq \mathbb{Z}^{d} $. The elements of $ l \, \in L $ satisfies  $ \sum_{i=1}^{n} l_{i}m_{i}=0 $, for the details of the proof, see ref.

\begin{definition}\label{Definition 2.1.3.} Let be $ L \, \subseteq \, \mathbb{Z}^{n} $, a sub-lattice.\\
\textbf{(a).} The ideal $ I_{L} = \langle \,  t^{\alpha} - t^{\beta} \vert \alpha - \beta \, \in L, \, \alpha, \, \beta \in \, \mathbb{Z}_{+}^{n}  \, \rangle$, is called a \textbf{lattice ideal}.\\
\textbf{(b).} A prime lattice ideal is called a \textbf{toric ideal.}\\\end{definition}
\textbf{Proposition 3.1.}\label{Proposition 2.1.} \cite{DCoxLittleSchenck2010}. An ideal $ I \, \subseteq \, \mathbb{C}[t_{1},...,t_{n}] $ is toric if and only if it is prime and it is generated by binomials.

\subsection{Toric Morphisms.}

We introduce in this section the concept of toric morphism. We will use this object in the conception of toric resolutions for singularities on a non-regular or singular variety.\\

\begin{definition}\label{Definition 2.3.0.} \cite{GEwald1993}. Let  $ \Phi :\mathbb{C}^{k} \longrightarrow \Phi(\mathbb{C}^{k}) $ be a monomial map, i.e., each non-zero component of $ \Phi $ is a monomial with coordinates in $ \mathbb{C}^{k} $, let $ X_{\sigma} \hookrightarrow  \mathbb{C}^{k}$ and $ X_{\sigma'} \hookrightarrow  \mathbb{C}^{m}$  be inclusions of toric affine varieties. If $ \Phi (X_{\sigma}) \, \subset \, X_{\sigma'} $, then $ \varphi  := \Phi \vert _{X_{\sigma}}$ is called a \textbf{toric affine morphism} of $ X_{\sigma} $  to $ X_{\sigma'} $. If $ \varphi $ is  bijective and its  inverse map $ \varphi^{-1} :X_{\sigma'} \longrightarrow X_{\sigma} $ is also a toric morphism, then  $ \varphi $ is called an \textbf{affine toric isomorphism} and it is denoted by $ X_{\sigma} \simeq X_{\sigma'} $.
\end{definition}

%\begin{definition}\label{Definition 2.3.3.} 
Recall  that a  \textbf{complex projective n-space $ \mathbb{CP}^{n} $} is the space of class of equivalence of pairs of points such that it consists of lines on $ \mathbb{CP}^{n}=\mathbb{C}^{n+1} / \sim  $. The relationship between points $ \sim $ is of the following manner. Given any vector $ v:= (\eta_{0},...,\eta_{n}) $ it defines a line $ \mathbb{C}*v $ and two of said vectors $ v \sim v' \in \mathbb{C}^{n+1}\setminus \lbrace 0 \rbrace $ define the same line if and only if, one is a multiple scalar of the other.\\

%\end{definition}\

We say, that when there exists an inverse toric morphism in any point $ p \, \in \, X_{\sigma'} $, this toric morphism defines a \textbf{toric resolution} on $ p $ or toric blow up for distinguishing it from a blowing up usual, and these maps will coincide. Taking $ X_{\sigma'} \simeq \mathbb{C}^{2} \times \mathbb{CP}^{2} $, we will see an explicit toric resolution in the k-dimensional complex projective-space $ \mathbb{CP}^{k-1}  $ in the last theorem \ref{Theorem of toric modification} which uses the Hironaka-Atiyah's theorem. Both of these theorems, are important tools for toric resolutions of genetic-metabolic networks.

\begin{lemma}\label{Lemma 3.1.0}\cite{GEwald1993}. Let $ \sigma $  be a lattice cone and set $ \tau \, \preceq \, \sigma  $ a face. Then there exists a natural identification, of toric varieties, given by the toric morphism,
\begin{center}
$  X_{\tau^{\vee}} \: \simeq \: X_{\sigma^{\vee}} \setminus \lbrace  u_{k}= 0 \rbrace$.
\end{center}
where $ u_{k} $ is the last generator of the representation of the coordinate ring associated to $ X_{\sigma^{\vee}} $.\end{lemma}
\begin{definition} \label{Definition 2.3.3.1.} The isomorphism,
\begin{center}
$ \psi_{\sigma, \sigma'} : \,  X_{\sigma^{\vee}} \setminus \lbrace  u_{k}= 0 \rbrace \longrightarrow  X_{\sigma'^{\vee}} \setminus \lbrace  v_{l}= 0 \rbrace $.
\end{center}
is called \textbf{gluing morphism}, which glues the varieties $ X_{\sigma^{\vee}} $ and $ X_{\sigma'^{\vee}} $ in the variety $ X_{\tau^{\vee}} $.\end{definition}

%\textbf{Proposition 2.2.}\label{Proposition 2.2.} Every toric morphism $ \varphi : X_{\sigma} \longrightarrow X_{\sigma'}$ determines a monomial homomorphism $ \varphi^{*} : R_{\sigma'} \longrightarrow R_{\sigma}  $ and reciprocally, for details of the proof, see \cite{GEwald1993}.

%\begin{definition}\label{Definition 2.3.1.} For two lattice cones $ \sigma \, \subset \mathbb{R}^{n}= lin(\sigma) $ and $ \sigma' \, \subset  \mathbb{R}^{m}= lin(\sigma') $, we say that $ \sigma $ and $ \sigma' $ are isomorphic  and denote  $ \sigma \, \simeq \, \sigma'$, if $ m= n $ and  there exists an uni-modular transformation $ L : \mathbb{R}^{n} \longrightarrow \mathbb{R}^{n} $ such that $ L(\sigma')=\sigma $. The monoids $ \sigma \cap \mathbb{Z}^{n} $ and $ \sigma' \cap \mathbb{Z}^{n} $ are isomorphic also.\end{definition}

%\begin{definition}\label{Definition 2.3.2.} Given $ R_{\sigma} $ and $ R_{\sigma'} $ two $ \mathbb{C}-$ algebras, there is a \textbf{monomial isomorphic},    $ R_{\sigma} \simeq R_{\sigma'} $, if there exists an invertible monomial homomorphism $ R_{\sigma} \longleftrightarrow R_{\sigma'} $\end{definition}
With the following theorem, we aim to highlight the joint the importance of the concepts; lattice polytope, polynomial $ \mathbb{C}- $algebras and toric variety. And of the morphisms associated to them, for a novel representation and study of genetic-metabolic networks. 

%In our propose in the following the explication of the result posterior and to denote the importance in genetic-metabolic networks, and it will be the great importance for the analysis in steady-state and dynamical of networks, even though our works has limited for a steady-state study only, and posteriori we will try the dynamical case. 
 ′
\begin{theorem}\label{Theorem 2.3.0.} \cite{MPCastilloOscar2017} and \cite{GEwald1993}. Set $ \sigma \, \subset \mathbb{R}^{n}= lin(\sigma) $ and $ \sigma' \, \subset \mathbb{R}^{m}= lin(\sigma') $, then  the following conditions are equivalent: 
\begin{center}
\textbf{(a).} $ \sigma \, \simeq \, \sigma' $  {} {} {}  \textbf{(b).} $ R_{\sigma} \simeq R_{\sigma'} $  {} {} {} \textbf{(c).} $ X_{\sigma} \simeq X_{\sigma'} $.
\end{center}
\end{theorem}
%\begin{proof} 
The implications a) $ \Rightarrow $ b)$ \Rightarrow $ c) are proven by means of the following diagram and, previously, we prove that it is commutative. 
\begin{center}

$ \qquad \qquad \sigma \quad \longrightarrow \quad R_{\sigma}  \quad \hookrightarrow\quad X_{\sigma}=$\textbf{Spec}$(R_{\sigma})$\\
$ \qquad \qquad \downarrow \uparrow  L^{-1}  \qquad  $ $ \downarrow \uparrow   \psi^{-1} \qquad \quad \quad\quad\quad $ $\downarrow \uparrow \varphi^{-1} $\\
$ \qquad \qquad \sigma^{'} \quad \longrightarrow \quad R_{\sigma^{'}} \quad \hookrightarrow\quad X_{\sigma^{'}}=$\textbf{Spec}$(R_{\sigma^{'}})$.
\end{center} ́
In the next section, we will point out where the principal concepts of systems biology are localized in Savageau, taking as a base the previous scheme.

%In this scheme, we will start by pointing and focus where are localized the principal concepts about systems biology in the approach of S-Systems defined by Savageau.\\
In this work, we will combine the algebraic geometry concepts defined in the diagram above, with regulatory-metabolic networks defined as sets of S-systems. We represent an S-system as a set of non-linear algebraic equations. These equations represent the change of concentrations of some species in the genetic-metabolic network over time. In general, these equations can be written in the \textbf{formalism of power law}, where we invite you to seen in the beginning of section four.\\
%Firstly we represent a S-Systems by means of a set of non-linear algebraic equations, and they are a representation of change of concentrations of some specie on the gene-metabolic network, the set is given by a expression of type, 
%\begin{center}
%\[ \dot{X}_{i}= \sum_{k=1}^{p_{i}} \alpha_{ik} \prod_{j=1}^{m}X_{j}^{g_{ijk}} - \sum_{k=1}^{q_{i}} \beta_{ik}\prod_{j=1}^{m} X_{j}^{h_{ijk}},\: i=1,...,n.\] 
%\end{center}
Where $\dot{X}_{i}  $ denotes the change of concentration of i-esim specie. And in agreement with our scheme above $\dot{X}_{i}$ is an element of the coordinate ring, i.e; $\dot{X}_{i} \, \in \, R_{\sigma} $ the S-Systems in general live in this ring. But in this polynomial ring. In general, the S-Systems exist in this ring. Any polynomial ring, including the particular ring previously described, can be singular. We can see in the following demonstration how any S-system is singular., i.e.
\begin{center}
\[ \dfrac{\partial \dot{X}_{i}}{ \partial X_{j}}(0,...0)= 0,\] 
\end{center}
without loss of generality for the zero concentration vector $  \bar{0}=(0,...,0) \, \in \, \mathbb{R}^{m}$, that represents the moment in which the concentration for any species is 0. In this lattice vector is clear to see the singularities $ \forall \, i=1,...,n $, it is always possible to prove by a translation of zero vector, for any point $ p \, \in \, \mathbb{R}^{m} $. This is an important fact for the dynamical solution in general. This problem combined with the lack of information about kinetic parameters and orders of reactions imply that for some values, some of the studied polynomials can tend to infinity. This is translated into a bad computational behaviour and a  non-obvious, or impossible biological situation. To solve this S-system, we translate this polynomial that belongs to the ring $ R_{\sigma} $.\\
So we have the polynomial $\dot{X}_{i} \, \in \, R_{\sigma} $, which is the singular case.
And now, to solve and characterize the S-Systems in a way most suitable. We translate this polynomial into a lattice-polytope utilizing the concept of Newton polyhedron, which consists of the lattice vectors of kinetic orders associated to an exponent-space. Formally we call it, the support of the polynomial, and we denote by $supp( \dot{X}_{i} )$, which in most of our example of S-Systems consist of two sums of binomials. For example, lets solve a genetic-metabolic network of five nodes. X is the concentration vector of all the nodes, so  $ X=(X_{1}, X_{2}, X_{3}, X_{4}, X_{5}) \, \in \, \mathbb{R}^{5} $. Now, let us consider only one equation of the system. \\
\begin{center}
   \[ \dot{X}_{3}=\alpha_{12} X_{1}^{g_{13}} X_{2}^{g_{12}} -  X_{3}^{h_{33}}.\] 
\end{center} 
Here $ \dot{X}_{3} \, \in \, R_{\sigma} $ since that the support associated to it, is 
\begin{center}
$ supp( \dot{X}_{3} )=\lbrace (g_{13}, g_{12},0,0,0), \, (0, 0, h_{33}, 0, 0) \, \in \, \mathbb{Z}^{5} \,\, \rbrace $,
\end{center}
at the same time the lattice vectors generate the cone $ \sigma $ and we denote $ \sigma=Con((g_{13}, g_{12},0,0,0), (0, 0, h_{33}, 0, 0)) $. Now, we have translated the problem  into a lattice cone, such that $supp( \dot{X}_{3} ) \, \in \, \sigma  $, the geometrical properties of lattice cone are limited algebraically, but we provide this linear space with the characteristics of a monoid by intersecting it with the lattice group $ \mathbb{Z}^{5}  $, so $ M=\sigma \cap \mathbb{Z}^{5} $.\\
Next, we will explain the construction of the toric variety associated to an S-system, a very interesting object. We take the two lattice vectors of support, we have $ (g_{13}, g_{12},0,0,0)=g_{13}e_{1} + g_{12}e_{2} $ and $ (0, 0, h_{33}, 0, 0)=h_{33}e_{3} $, we associate the monomials to this lattice vector as: $  X^{g_{13}e_{1} + g_{12}e_{2}}=X_{1}^{g_{13}} X_{2}^{g_{12}}$ and $ X^{h_{33}} =X_{3}^{h_{33}}$. The toric variety associated to this terms are the primes monomials, which are denoted by $ X_{\sigma}= \lbrace X_{1}^{g_{13}}, X_{2}^{g_{12}}, X_{3}^{h_{33}} \rbrace$. This equivalent to say that this set is the algebraic affine scheme associated to the polynomial coordinate ring generated for the monomials, we write that as: $ X_{\sigma^{\vee}} =$\textbf{Spec}$  \mathbb{C}[X_{1}^{g_{13}}, X_{2}^{g_{12}}, X_{3}^{h_{33}}]= $\textbf{Spec}$ (R_{\sigma}) $.\\

The next figure summarizes the facts mentioned previously. It is a schematic representation of the morphism associated to the commutative diagram shown in Theorem 3.3. In the Figure 1.0 we can see the problems generated when there are singularities in any of three different levels. In the middle of the figure the set of equations of any S-system associated to a gene-metabolic network is represented. This set belongs to a finitely generated $ \mathbb{C} $-algebra which has some singular polynomials. One can take the prime polynomials associated to these singular polynomials to construct an irregular toric variety or not complete variety. Also, we can construct the irregular lattice polytope (left side of the figure) from the support space of these singular polynomials, also called exponent-space or Newton polyhedron. The lattice vectors associated as vertex of these irregular lattice polytope, are represented with a matrix array in row, this representation induces a linear transformation $ L $, in the dictionary of linear transformations. When the lattice polytope be regular then the matrix associated $ L_{A} $ to this, will have the property of uni-modularity, i.e., \textbf{det}$ L_{A}=\pm 1 $, and the linear transformation $ L $ now is named as uni-modular transformation. It is work of us to transform  the irregular lattice polytope to a regular lattice polytope by means of uni-modular transformation and these are done by means of Hilbert basis generated from the set of vertex of irregular polytope.
%To continuation we will resume in the figure below, the facts mentioned in the previous example, an schematic representation of the morphism associated to the commutative digram above, in these maps we have the problems that represent to have singularities with the S-Systems in three different levels: first, the set of equations of any S-system associated to a gene-metabolic network, belongs to a finitely generated $ \mathbb{C} $-algebra, and this is the representation in middle on the figure; second, these are singular polynomials belonging to this algebra, and now we can represent the same issue in one way such that, these singular polynomials are mapped into an irregular toric variety or not complete variety, it is constructed taking the prime polynomials associated of initial set in middle; and third, in this case the representation is the irregular lattice-polytope of left side in the diagram, and it is constructed of the support space of singular polynomial in middle, also is called the exponent-space or Newton polyhedron; since that is the geometrical realization of a polytope, where each vertex coordinate is yields from each one of the monomials  in an only equation of each singular polynomial associated to the S-System associated in middle on the figure, and this polytope is irregular since that any square sub-matrix associated has determinant $ \neq \pm 1 $ (uni-modular transformation), is to say this set of vertex is not isomorphic to an octant of Euclidian space, necessary condition for regular or non-singular polytopes. \\ 

\begin{figure}[htbp]
\centering
\makebox{\includegraphics[height=2.5in]{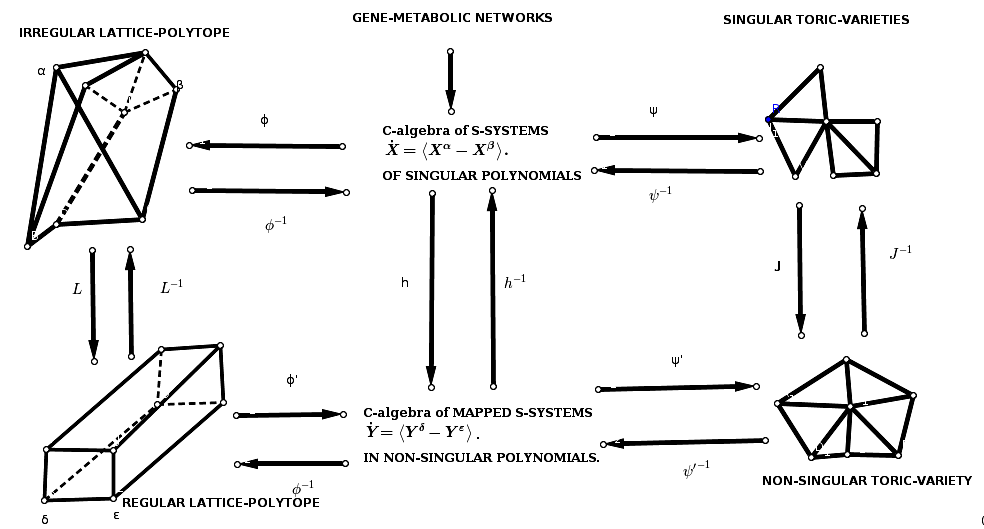}}
\caption{\label{fig_simvarex} Three different representations of singular S-Systems and three representations for non-singular S-systems.}
\end{figure}
The problem of having singularities is that they increase the complexity of the system, making it more difficult to solve systems with a large number of equations, such as large-scale genetic-metabolic networks. This problem can be tackled using tools from algebraic geometry. This can be solved by mapping any of these objects: irregular lattice-polytope, singular S-systems polynomials, or irregular toric varieties, into regular or non-singular forms.\\
 This solution is represented in the next figure, in which each irregular object is mapped to its regular or non-singular counterpart.
 
 %The solution to this issue is represented in the figure below, where each object is mapped to a regular or no-singular object corresponding to each irregular object of the previous figure, as we see.\\

%In the next one, we see that the problematic to have singularities into of any S-System, it which difficult the computational solutions of a huge number of equations and therefore an inadequate approach for modelling of large-scale gene-metabolic networks, this problematic is homework to the algebraic geometry and can be solved, mapping any of these objects: irregular lattice-polytope, singular S-systems polynomials, or irregular toric varieties, into regular or non-singular forms.\\
 %The solution to this issue is represented in the figure below, where each object is mapped to a regular or no-singular object corresponding to each irregular object of the previous figure, as we see.\\
%\begin{figure}[htbp]
%\centering
%\makebox{\includegraphics[height=2.0in]{Regular_variety.png}}
%\caption{\label{fig_simvarex} Three different representations of singular S-Systems with high computational complexity.}
%\end{figure}

The principal methodology developed in our work transforms irregular lattice polytopes into regular or non-singular polytopes. This will allows us to transform a set of singular polynomials, that represent S-systems into a set of non singular polynomials of S-Systems. This will open the possibility to model large-scale metabolic networks.\\
In our work, we study the irregular lattice-polytope (Figure 1, top left panel). We transform the irregular lattice cone into a monoid by intersecting with $ \mathbb{Z}^{n} $, which can be written as $ M= \sigma\cap \mathbb{Z}^{n} $. By applying the Gordan's lemma to the monoid we can conclude that this is finitely generated, that is to say there exist a set of generators such that they generate all M. This set of generators is minimal and is named the Hilbert basis. The elements of Hilbert basis are lattice vectors and form a regular or non-singular
lattice polytope (Figure 1, bottom left panel). By arranging these vectors into a matrix, we induced a uni-modular transformation, $ L $. i.e; \textbf{det}$ L_{A}=\pm 1 $. This transformation and its inverse transformation $ L^{-1} $  are used to map irregular lattice-polytopes into regular lattice-polytopes (Figure 1, arrows between top and bottom panel).\\
This mapping is the  algorithmic basis to computationally transform irregular lattice-polytopes into regular ones.

%And this solution consists in to study the irregular lattice-polytope in our case we see the lattice cone, as a monoid, we have a monoid  simply intersecting with $ \mathbb{Z}^{n} $ and the irregular lattice con, this is seen as, $ M= \sigma\cap \mathbb{Z}^{n} $, by the Gordan's lemma the monoids are finitely generated, is to say there exist a set of generators such that generate all $ M $. This set of generators is minimal and is named the \textbf{Hilbert basis}, the elements of Hilbert basis are lattice vectors and form a regular or non-singular lattice polytope, the matrix array of these vectors induced a unimodular transformation, i.e; \textbf{det}$ = \pm 1 $ in the figure below is denoted by the linear transformation $ L $ and inverse transformation $ L^{-1} $. With the map $ L $ we can computationally to map irregular lattice-polytope into regular lattice-polytope the basis is an important algorithmic tool for this homework as we can see.\\

%\begin{definition}\label{Definition 2.3.3.} Recall  that a  \textbf{complex projective n-space $ \mathbb{C}P^{n} $} is the space of class of equivalence of pairs of points such that it consists of lines on $ \mathbb{C}P^{n}=\mathbb{C}^{n+1} / \sim  $. The relationship between points $ \sim $ is of the following manner, given any vector $ v:= (\eta_{0},...,\eta_{n}) $ it defines a line $ \mathbb{C}*v $ and two of said vectors $ v \sim v' \in \mathbb{C}^{n+1}\setminus \lbrace 0 \rbrace $ define the same line if and only if, one is a scalar multiple of the other.\end{definition}\

\begin{theorem} \label{theorem 4.} (\textbf{\textit{Hironaka-Atiyah}}), \cite{MFAtiyah1970}. Let $ g $ be a real analytical function $ (\neq 0) $ in a neighbourhood of $ p_{0} \, \in \, \mathbb{R}^{n} $. Then there exist an open set $\omega \, \in \, W \subset \, \mathbb{R}^{n} $, a real analytical variety $ W_{0} $ and a proper analytical map $ \varphi :W_{0}$ to $W$ such that:\\
i). $ \varphi :(W_{0}-A_{0}) \longrightarrow (W-A) $ is an isomorphism, where  $ A =g^{-1}(p_{0}) $ and $ A_{0}=\varphi^{-1}(A)=(g(\varphi))^{-1}(p_{0}) $;\\
ii). For each $ \omega  \, \in W_{0} $, there are local analytical coordinates  $ (\omega_{1},....,\omega_{n}) $ such that there exists a resolution map, we write $ g(\varphi(\omega))=\pm h_{1,...,n} \omega_{1}^{r_{1}}*\omega_{2}^{r_{2}}*...*\omega_{n}^{r_{n}} $, where $ r_{1},....,r_{n} $ are  non negative integers and $ h_{1,...,n} $ is an invertible analytic function; see ref.
\end{theorem}

The theorem is a modified version of the theorem "resolution of singularities" and is used for analytical varieties. The original theorem was proved by Hironaka, \cite{HHironaka1964}.\\

%The previous theorem is a version for analytical varieties of theorem of resolution of singularities proved by Hironaka in algebraic geometry, see ref. \cite{HHironaka1964}.\\

In the following we enunciate a general result for  non regular or singular varieties. 
\begin{theorem}\label{Theorem of toric modification} \cite{GEwald1993}. Let $ X_{\sum} $ be a regular toric variety, and let $ X_{\sum_{0}} $ be a toric invariant sub variety defined by the star
$ st(\sigma,\sum) \simeq  \sum_{0}  $ of
$ \sigma $ into $ \sum $; 
$ 1 < k:=dim \sigma \leq n $.\\
Then under toric blow up $ \psi^{-1}_{\sigma} $, any point $ x \in X_{\sum_{0}} $ is  substituted by a k-dimensional \textbf{(k-1) projective space.}\\
%(b) The blow down $ \psi_{\sigma} $ is a toric morphism which is bijective in the outside of $ \psi^{-1}_{\sigma} $.\\
See ref. for the proof of this fact.
\end{theorem}
This theorem says that if we have a singular variety in any point $ p \, \in \,\mathbb{R}^{m} $, i.e. the condition is true. So, the variety $ X_{\sigma_{0}} $  built from its support polynomial $ \dot{X}_{i}=0 $, is a non regular toric variety, which implies that it has singularities. By Hironaka's theorem  \cite{HHironaka1964}, there exists a morphism called resolution or toric blow-up, which yields a resolution of these singularities. Using this morphism the variety $ X_{\sigma_{0}} $ is mapped  into a regular toric variety $ X_{\sigma}$ or non singular variety. It is possible to show that the transformation of coordinates consists in mapping each singularity into a complex n-dimensional (n-1)-projective space as pointed in the last theorem \label{Theorem of toric modification}, i.e. we take the coordinates of singular point and we substitute them by a dimensional (k-1)-projective space, where $ k $ depends on the dimension, see \cite{GEwald1993}. The parameter space is of special interest for the study of the dynamics of gene regulation, as we will see.

\section{Biochemical Machines: Toric resolution in genetic-metabolic networks.}
In this section we introduce biochemical systems particularly genetic-metabolic networks modelled using the power law formalism of enzyme kinetics, defined and widely studied by M. Savageau, \cite{MASavageau1976}, \cite{MASavageauVoit1987}, \cite{MASavageau2001}. Subsequently, we prove that they are toric varieties.\\\\

\textbf{Formalism of power law for biochemical systems.}\\
The power law formalism is a particular representation of enzyme kinetics, formally a Taylor's expansion in several variables capable of describing biochemical reactions of metabolic and gene regulatory mechanisms, formulated as rational functions produced by the balance of mass equations.
\begin{center}
\[ \dot{X}_{i}= \sum_{k=1}^{p_{i}} \alpha_{ik} \prod_{j=1}^{m}X_{j}^{g_{ijk}} - \sum_{k=1}^{q_{i}} \beta_{ik}\prod_{j=1}^{m} X_{j}^{h_{ijk}},\: i=1,...,n.\] 
\end{center}
This expression represents the basic principle of mass conservation, called \textbf{general mass action (GMA)}, in the form of the power law formalism. Each term is associated to elements of a network of reactions, where  $\dot{X}_{i} $ represents a derivative of $ X_{i} $ over time t. This expression represents the rate law of degradation or production of any $ i $-esim species $ X_{i} $ (gene, mRNA, proteins or metabolites) in the network. This variation in time describe the flux of biomass throughout all elements in any biochemical reaction and genetic network. The first sum is that of the products that contribute to the production to the species $ X_{i} $ and the terms of the second sum are elements of consumption due to the degradation or dilution of $ X_{i} $, \cite{MASavageau1976}, \cite{MASavageau2001}.\\
\begin{definition}\label{Definition 6.0.}(\textit{\textbf{Dominant terms}}). A dominant term is defined
as the largest term of a given sign for an equation of the GMA-system (equation defined above); the dominant terms with positive and negative signs are the \textbf{dominant positive term} and the \textbf{dominant negative term}, respectively, see ref. \cite{MASavageau1976}.\end{definition} \

%\textbf{Definition 6.0. (Dominant terms).} For the expression defined above, for all $ i=1,...,n $, the largest term in the sum is defined as the dominant term; if this term corresponds to a term in the positive sign sum, it is defined as the dominant positive term, and similarly for the negative sum, it is defined as the dominant negative term.\\

\begin{definition}\label{Definition 6.1.} (\textit{\textbf{S-systems}}). Using the concept defined above, we can have a finite number of combinations of dominant terms, exactly  \[ \prod_{j=1}^{m} p_{j}q_{j} ; \]  partitions of the space of concentrations $(X_{1},...,X_{n}) $, where each partition is a dominant sub-system and it is associated with a particular system of equations, called \textbf{S-systems}, M. Savageau, \cite{MASavageauVoit1987}, \cite{MASavageau1976}, \cite{LomnitzSavageau2013}, \cite{LomnitzSavageau2014}, \cite{LomnitzSavageau2015}; as shown below
\[ \dot{X}_{i}=  \alpha_{ip_{i}} \prod_{j=1}^{m} X_{j}^{g_{ip_{i}}} -  \beta_{iq_{i}} \prod_{j=1}^{m}X_{j}^{h_{iq_{i}}}, \, i=1,...,n. \].\\
%\[ 0=  \alpha_{ip_{t+1}} \prod_{j=1}^{m}X_{j}^{g_{ip_{t+1}}} -  \beta_{iq_{t+1}} \prod_{j=1}^{m}X_{j}^{h_{iq_{t+1}}}, \, t=0,...,m , \]  
for details, see M. Savageau \cite{MASavageauVoit1987}.\end{definition}\
Note in this last definition that the sums were neglected, at difference to the formalism power law formalism complete. An S-System only consist of dominant terms.
\begin{definition}\label{Definition 6.2.} (\textit{\textbf{Phenotypic Polytope}}). A phenotype is defined as a set of concentrations or fluxes corresponding to a valid combination of dominant terms. They define boundary conditions obtained from the solution of S-systems under steady-state, in logarithmic-space. Each boundary condition defines an m-dimensional half-space and the intersection of these half-spaces yields an m-dimensional \textbf{phenotypic polytope}. \end{definition}\
The phenotypic polytope associated with the S-Systems can be considered to be a realization of a lattice polytope, taking the support of each S-systems related to the kinetic orders of the metabolic reactions.\\
Using the previous definitions, in t­he following Theorem we prove that the S-Systems are toric varieties.
%\subsection*{Toric Embedding of S-systems}.
\begin{theorem}\label{Theorem 1.} (\textit{\textbf{Toric Embedding of S-systems}})
 Let $ \dot{X}_{i} $ be an S-system, with initial conditions $ X_{i}(0)=X_{io} \,\in\, \mathbb{R}$, for $ i=1,...,n $.\\
 %given by:
%\begin{center}
%\[ \dot{X}_{i}=\alpha_{i} \prod_{j=1}^{m}X_{j}^{g_{ij}} -  \beta_{i}\prod_{j=1}^{m}X_{j}^{h_{ij}} ; \:  X_{i}(0)=X_{i0}. \]
%\end{center}
Then the S-system $ \dot{X}_{i}$, for all $ i=1,2,...,n $, is generated by \textbf{toric ideals} $ I_{H}$ and as a consequence, each $ \sigma_{i} \subseteq \, \mathbb{Z}^{n} $ \textbf{phenotypic polytope} is associated to an affine toric variety, namely,
\begin{center}
$ X_{\sigma^{\vee}_{i}} =$\textbf{Spec} $( R_{\sigma_{i}})$, $ \: \forall \, i=1,2,...,n,  j=1,...,m $.\\
 
\end{center}
With coordinate ring $ R_{\sigma_{i}} $ associated to the monomial $ \mathbb{C}- $algebra, $ \mathbb{C}[X_{j}^{g_{ij}}, X_{j}^{h_{ij}}] $, and it is called a \textbf{toric phenotypic variety} or simply \textbf{phenotypic variety}.
\end{theorem}
\begin{proof}
 We prove the result by construction, given the following polynomial  $ \dot{X}_{i} =\alpha_{i} \prod_{j=1}^{m}X_{j}^{g_{ij}} -  \beta_{i}\prod_{j=1}^{m}X_{j}^{h_{ij}} $, for some $ i $ which is an S-system, where $\dot{X}_{i} \, \in \, \mathbb{C}[X_{j}^{g_{ij}}, X_{j}^{h_{ij}}] $. It is clear to see that the monomial $\mathbb{C} -$algebra is generated by the monomials $ X_{j}^{g_{ij}} $ and $ X_{j}^{h_{ij}} $; and  always an S-system is defined by two positive condition lattice vectors $ a_{i}=(g_{i1},...,g_{im}), \: b_{i}=(h_{i1},...,h_{im}) \: \in \: \mathbb{Z}_{+}^{m} $. Allow these vectors of kinetic orders which are associated to each binomial of each equation of an S-system consist of positive terms, otherwise,  $ g_{ij} $ and $ h_{ij} $ could be rational numbers or they could be negative integers. In steady-state, i.e., $ \dot{X}_{i}=0 $, one can always multiply by units of the ring $ X^{\pm r} $ and $ X^{\pm s} $ the polynomial  $ \dot{X}_{i} =0$, with $ r, s \: \in \, \mathbb{Z} $, such that, $ rsa_{i}, \: rsb_{i}\: \in \: \mathbb{Z}_{+}^{m} $ or equivalently once that we prove the S-Systems are toric, we see that exists an invariance for action torus ($ T \times X_{\sigma^{\vee}} \longrightarrow X_{\sigma^{\vee}} $, given by $ (t,x)\longmapsto t*x=(t^{a_{1}}*x_{1},...,t^{a_{k}}*x_{k})$ with $t^{a_{i}}\, \in\, \mathbb{C}^{*}  $).\\
 
Now set $ \sigma $ to be the lattice cone generated by the two rays $ a_{i} $ and $ b_{i} $. We will prove that its dual lattice cone is strongly convex, i.e., it complies with the property $ \sigma^{\vee} \cap (-\sigma^{\vee})=\lbrace 0 \rbrace $, so that we can apply the Hilbert basis's lemma. First, we note that the lattice vector $ a_{i}- b_{i} \,\in\, \sigma$ if and only if $ a_{i} \preceq b_{i}  $ where $ \preceq $ is a partial order induced by the inclusion on the faces of the lattice cone, and therefore the zero lattice vector is a face, $ 0\preceq \sigma $. This assumption is true, if and only if $\sigma^{\vee}  $ is full dimensional, i.e., that it contains bases as vector space to generate an octant of n-dimensional Euclidian-space; and if and only if $\sigma^{\vee}  $ is strongly convex. Note that, all these facts are standard results of convex geometry.\\
%The assumption is also true, if $ a_{i} \preceq b_{i} $, if and only if $ \sigma $ is strongly convex, i.e., $\sigma \cap (-\sigma)=\lbrace 0 \rbrace   $, so $ \sigma $ is strongly convex, if $ \sigma $ also implies that $\sigma^{\vee}  $ is full dimensional, i.e., that it contains bases as vector space to generate an octant of n-dimensional Euclidian-space. Therefore dim$ \sigma^{\vee}=n $ implying that $\sigma^{\vee}  $ is strongly convex, then $\sigma^{\vee} \cap (-\sigma^{\vee})=\lbrace 0 \rbrace $. Note that, all these facts are standard results of convex geometry.\\
%In order to apply Hilbert basis's lemma, we construct the dual lattice cone $ \sigma^{\vee}_{i} \, \subseteq \, \mathbb{Z}^{m} $ using the implications above, since it is strongly convex if and only if it is a partial ordered lattice $ a'_{i} \preceq b'_{i}  $ which also implies that $ a'_{i} - b'_{i} \, \in \, \sigma^{\vee}_{i};  \, a_{i}', b'_{i} \in \, \mathbb{Z}^{m} $. We define the monoid $ S_{\sigma_{i}}= \sigma_{i}^{\vee} \cap \mathbb{Z}^{m} $; this monoid is finitely generated (Gordan's lemma), and therefore since that $\sigma^{\vee}_{i}  $ is strongly convex. Applying Hilbert basis' lemma, $ S_{\sigma_{i}} $ contains a finite minimal set of generators.
Since $ \sigma^{\vee}_{i} \, \subseteq \, \mathbb{Z}^{m} $ is strongly convex, we can use the Hilbert basis theorem to obtain a finite minimal set of generators for the monoid $ S_{\sigma_{i}}= \sigma_{i}^{\vee} \cap \mathbb{Z}^{m} $. We order them in a row matrix array  $ H=\lbrace h_{ij} \rbrace $ (Hilbert basis). The ideals associated to them are: 
\begin{center}
$ I_{H}=< X^{a'_{i}}- X^{b'_{i}} \, \vert \, (a'_{i} - b'_{i}) \, \in \, ker(H) > $,
\end{center} 
These ideals are primes and generate $ \dot{X}_{i} $, to show that are primes, we chose any binomial, and is sufficient to show that it is prime, suppose that it is not, so it is factorized at least for two elements in the $ \mathbb{C}-$algebra, i.e., $x^{a}-x^{b}=f*g  $, we construct the Newton polyhedron, taking $supp(x^{a}-x^{b})=supp(f*g)=supp(f)+supp(g)$ (property of Minkowski sum) \cite{BSturmfels1996}, therefore the elements that generate the lattice cones $supp(f)$ and $ supp(g) $, also generate the lattice cone $supp(x^{a}-x^{b})=Con(a,b)$, but $ a,b \,\in\,H$ are minimal elements, from here the contradiction, therefore $x^{a}-x^{b}$ is prime. Now the application of the definition \ref{Definition 2.1.3.} implies that they are lattice ideals, in agreement with the hypothesis of the proposition 3.0. We conclude that they also are toric ideals. Now, we construct the regular lattice cone $ \sigma^{\vee}_{iH} $ associated to these Hilbert basis, where the regularity is provided  by the fact that, det$ (H)=\pm 1 $, this process is valid for each $ i=1,...,n $. With these facts and using the theorem \ref{Theorem 2.3.0.}, we construct the coordinate ring $ R_{\sigma^{\vee}_{Hi}}= \lbrace \dot{X}_{Hi} \, \in \,   \mathbb{C}[\bar{X_{j}}^{g_{ij}},\bar{X_{j}}^{h_{ij}}] \vert \: supp(\dot{X}_{Hi}) \subset  \sigma^{\vee}_{Hi} \rbrace $, with lattice cone $ \sigma^{\vee}_{Hi}= Con (H) $ generated by the Hilbert basis, where we made a change of coordinates of the binomials $ \dot{X}_{i} $ in the system provided by the Hilbert basis, and we denote them now by $\dot{X}_{iH}$. This change of coordinates is always possible using the morphisms $ L_{H} $ and $ \chi $ defined in theorem \ref{Theorem 2.3.0.}. Also there exists a monomial homomorphism $ j_{\sigma^{\vee}_{Hi}} $ that give rise for all $ i$, into the affine algebraic scheme $ X_{\sigma^{\vee}_{Hi}}= $ \textbf{Spec} ($ R_{\sigma^{\vee}_{Hi}} $). This is a toric variety with monomial $\mathbb{C} -$algebra given by the isomorphisms classes $ \mathbb{C}[S_{\sigma^{\vee}_{Hi}}]=\mathbb{C}[X_{j}^{g_{ij}}, X_{j}^{h_{ij}}] / I_{H} \simeq  \mathbb{C}[\bar{X_{j}}^{g_{ij}},\bar{X_{j}}^{h_{ij}}]$ (toric re-parametrization), with new coordinate system given by the classes prime monomials $\bar{X_{j}}^{g_{ij}}  $ and $\bar{X_{j}}^{h_{ij}}  $. In this way, we have proven that  \textbf{S-systems are toric} or have an embedding in the algebraic torus, and $ X_{\sigma^{\vee}_{Hi}} $ is the associated \textbf{phenotypic variety}; with this fact we conclude the proof. \end{proof} 

S-systems are usually solved under steady-state conditions. In the following, we make a toric resolution in S-Systems without assuming steady-state. Theorem \ref{Theorem 2.} will enable the study of their dynamics.

\begin{theorem}\label{Theorem 2.}\textbf{(\textit{Toric resolution in S-systems})}. Let $ \dot{X}_{i} $ be an S-system.
%\begin{center}
%\[ \dot{X}_{i}=\alpha_{i} \prod_{j=1}^{m}X_{j}^{g_{ij}} -  \beta_{i}\prod_{j=1}^{m}X_{j}^{h_{ij}} ; \:  X_{i}(0)=X_{i0}. \]
%\end{center} 
%for all $ i=1,2,...,n, \: j=1,...,m $. 
Then there are maps of resolution $ \psi_{i} : X_{\sigma_{i}^{\vee}} \longrightarrow X_{\sigma_{i}^{' \vee}}$ for each $ i $ (toric morphisms) such that, they are a re-parametrization in the coordinate ring $ \mathbb{C}[\omega_{1},...,\omega_{m}] $ thus;
\begin{center}
$\dot{X}_{i}(\psi_{i}(\omega) )=h_{i} \omega_{1}^{\alpha_{1}}*.....*\omega^{\alpha_{m}}_{m}  $, with $ \omega_{i}(0)=\omega_{i0} $.
\end{center}
where the $h_{\alpha_{1},...,\alpha_{m}} $ are units of the ring, with $ \omega=(\omega_{1},...,\omega_{m}) \, \in  \, \mathbb{R}^{m}$ and  $ \alpha_{1} > 0,...., \alpha_{m} > 0 $, positive integers.\\
\end{theorem}
\begin{proof}
%The proof is given by construction and follows by applying  Theorems \ref{Theorem 2.3.0.}, and  Theorem of Hironaka-Atiyah  a finite number of times for each lattice cone $ \sigma_{i} $ up to reaching the toric resolution necessary for producing a polynomial $ \dot{X}_{i}(\psi_{i}(\omega) ) $ for some coordinate-space $\omega \, \in  \, \mathbb{R}^{m}  $; the lattice cones $ \sigma_{i}^{\vee} $ and $ (\sigma'_{i})^{\vee} $ are realized as follows, first: we take the lattice polytope for each S-system $ \dot{X}_{i} $, i.e., $ supp(\dot{X}_{i}) $, then, we compute the Hilbert basis $ H $ the necessary times to these associated lattice cones, obtaining the toric ideals $ I_{H} $ associated to these basis, and with the generators of the basis construct $ \sigma_{i}^{' \vee} $, and applying the Theorem \ref{Theorem 2.3.0.} on these lattice cones, we define the morphisms $ \varphi $ and $ \varphi^{-1} $. With these facts and in agreement with the Theorem of Hironaka-Atiyah, we re-parametrize the polynomial $ \dot{X}_{i} $ in the constructed coordinate ring as we can see, $  \mathbb{C}[\omega_{1},...,\omega_{m}] =\mathbb{C}[S_{\sigma'_{i}}]=\mathbb{C}[X_{j}^{g_{ij}}, X_{j}^{h_{ij}}] / I_{H} $, with $ S_{\sigma'_{i}} $ monoid associated to $ \sigma_{i}^{' \vee} $, and so we have the desired result, \textbf{q.e.d.} 
The proof is given by construction and by applying  Theorems 3.6, and the theorem of Hironaka-Atiyah, as we will see. First we take for some $ i $, with $ i=1,...,n $ an equation of an given S-system $\dot{X}_{i}  $, in agreement with  Theorem 3.6, let $\dot{X}_{i} \, \in \, R_{\sigma}  $.The definition of $R_{\sigma}  $ implies that:
\begin{center}
$supp( \dot{X}_{i})=\lbrace a_{i} \rbrace=\lbrace (g_{i1},...,g_{im}),(h_{i1},...,h_{im}) \rbrace \, \subseteq \, \sigma$;
\end{center}
we form the lattice cone $ \sigma=Con\left( (g_{i1},...,g_{im})-(h_{i1},...,h_{im})\right)  \,\subseteq \, \mathbb{Z}^{m} $ that consists, of the difference lattice vector, which implies the partial order $ \preceq $. i.e. Given $g_{i}=g_{i1},...,g_{im} $ and $ h_{i}=h_{i1},...,h_{im} $, set $ g \preceq h $ or vice versa, if and only if $ (g-h) \,\in\, \sigma $ for some lattice cone $ \sigma $, this construction of cones is important,  because agreement of definition 3.1 and Proposition 2.0, the difference $ g -h$ can be lifting to toric binomials $ (X^{g}- X^{h}) $, and we show it by the following process.\\

Given the lattice cone $ \sigma \,\subseteq \, \mathbb{Z}^{m} $ built as above. From the same Theorem 3.6, it follows the existence of the following morphisms, 
$ \chi_{i} \, : \sigma \, \longrightarrow  \, R_{\sigma}$, $ \chi^{-1}_{i} \, :R_{\sigma} \, \longrightarrow  \, \sigma $, $ L_{H} \, :\sigma \, \longrightarrow  \, \sigma'$, $ \chi_{i}' \, : \sigma' \, \longrightarrow  \, R_{\sigma'}$, $ \psi_{i} \, :R_{\sigma}  \, \longrightarrow  \, R_{\sigma'}$, $ \psi^{-1}_{i} \, :  R_{\sigma'} \, \longrightarrow  \,  R_{\sigma}$, where the $\psi_{i}  $ are morphisms of $ \mathbb{C} $-algebras.\\

%$ \chi_{i} \, :a_{i} \, \longrightarrow  \, X^{a_{i}}$;\\ 

Explicitly we have the maps, $ \chi_{i}(a_{i})=X^{a_{i}} \, \in \, R_{\sigma} $; $ \chi^{-1}_{i}(X^{a_{i}})=a_{i} \, \in \, \sigma $; $ L_{H}(a_{i})=h_{i} \, \in \, \sigma' $; $ \chi_{i}'(h_{i})=\omega^{h_{i}}=\omega_{1}^{h_{i1}}*...*\omega_{n}^{h_{im}} \, \in \, R_{\sigma'} $; $ \psi^{-1}_{i}(\omega)=\omega^{h_{i}}\, \in \, R_{\sigma} $, where $ a_{i}=(a_{i1},...,a_{im}) \, \in \, \mathbb{Z}^{m} $ is the condition lattice vector associated to $\dot{X}_{i}$, we also write $X^{a_{i}}=X_{1}^{a_{i1}}*...*X_{m}^{a_{im}} $. The map $ L_{H} $ represent the linear transformation of the matrix $ H=(h_{ij})$, this matrix is an array row, this consist of the Hilbert basis computed on the monoid associated to the non-regular lattice cone $ \sigma $, i.e, $ S_{\sigma}=\sigma^{\vee} \cap  \mathbb{Z}^{m}$. \\
Once we have computed the Hilbert basis on the monoid $ S_{\sigma} $, the matrix array induces an uni-modular transformation, i.e., det$H=\pm 1 $ (if $ m=n $, if $ m \neq n $ then we take squares sub-matrix), which is the condition for regular lattice cones, these basis $ H $ conform a new lattice cone $ \sigma' \,\subseteq \, \mathbb{Z}^{m} $. In this form we obtain the map of lattice cones $ L(\sigma)=\sigma' $, where $ \sigma' $ now represents a regular cone. From the commutativity property illustrated in the diagram of Theorem 3.6, we have the composition of maps,
\begin{center}
$ \psi^{-1}_{i} \, \circ \, \chi' = \chi \, \circ \, L_{H}^{-1}$.
\end{center}
from the equality of maps, we can obtain the following: 
\begin{center}
$ \psi^{-1}_{i}(\chi'(\omega))=\omega^{h_{i}}=\omega_{1}^{h_{i1}}*...*\omega_{n}^{h_{im}}=\chi( L_{H}^{-1}(h_{i}))=X^{L_{H}^{-1}(h_{i})} $. 
\end{center}
Therefore it is a new coordinate system induced from Hilbert basis, and we can re-parametrize each monomial of $ \dot{X}_{i} $, hence:
\begin{center}
$ \dot{X}_{i}(\psi^{-1}_{i}(\omega))=\alpha_{i}\omega_{1}^{h_{i1}}*...*\omega_{n}^{h_{im}}-
\beta_{i}\omega_{1}^{h'_{i1}}*...*\omega_{n}^{h'_{im}}   $.
\end{center}
some monomials can be factorized, because some integers can be $h_{ik} \geq h'_{il} $ or vice verse, for some $ k,l \,\in \, \lbrace 1,...m \rbrace$, i.e; if there exist terms with $ s_{ij}+p_{ij}=h_{ij} $ and $s_{ij}+q_{ij} = h'_{ij}$, then,
\begin{center}
$ \dot{X}_{i}(\psi^{-1}_{i}(\omega))=\omega_{1}^{s_{i1}}*...*\omega_{n}^{s_{ir}}(
\alpha_{i}\omega_{1}^{p_{i1}}*...*\omega_{n}^{p_{il}}-
\beta_{i}\omega_{1}^{q_{i1}}*...*\omega_{n}^{q_{im}} )  $.
\end{center}
The re-parametrization process is well defined. During this process we will apply Hironaka-Atiyah's theorem, given that the polynomial $ \dot{X}_{i} $ is an analytical function; if the polynomial $ \dot{X}_{i}(\psi^{-1}_{i}(\omega)) $ is a non singular polynomial the toric resolution
is completed; and we redefine $ f_{1,...,m}= \omega_{1}^{s_{i1}}*...*\omega_{n}^{s_{ir}}$ as an analytical invertible function, and  positive integers $ \alpha_{1}=s_{i1} >0,...,\alpha_{m}=s_{i1}>0 $, such that,
\begin{center}
$ \dot{X}_{i}(\psi^{-1}_{i}(\omega))= f_{1,...,m}\omega_{1}^{\alpha_{1}}*...*\omega_{n}^{\alpha_{m}}$.
\end{center}
Otherwise, if the resulting polynomial is singular, we repeat the process: We compute the Hilbert basis on the lattice cone of the support of the new polynomial $  \dot{X}_{i}(\psi^{-1}_{i}(\omega))$, until a non-singular polynomial is generated. The process is finite, and the maps are composed of finite number of times. The finiteness of the process is guaranteed by the Hironaka's theorem in its form of Principal Theorem I, which is a widely known theorem and it is not shown in this work, see \cite{HHironaka1964}. We also have that the morphism $(\psi^{-1}_{i}(\omega))  $ induce a toric morphism and yields a non singular toric variety, namely $ X_{\sigma'}= $\textbf{Spec}$(R_{\sigma})  $, whenever $\psi^{-1}  $ is non singular too. And with these facts, we conclude the proof. \end{proof}
%for latest we mention that the toric resolution with the parametrization $ (\psi^{-1}_{i}(\omega))  $ permit us to solve the dynamics of S-systems for genetic metabolic networks on the concentrations logarithmic-space.

\section{Applications of Toric S-Systems.}

%\subsection{Examples and Applications}\
%Prior to submission, authors who believe their manuscripts would benefit from professional editing are encouraged to use a language-editing service (see list at www.pnas.org/site/authors/language-editing.xhtml). PNAS does not take responsibility for or endorse these services, and their use has no bearing on acceptance of a manuscript for publication. 
%\textbf{Phenotypes in an oscillatory synthetic gene network.}\\\\
\textbf{Phenotypes in an oscillatory synthetic gene network.}\\

 In this example we will describe a gene network of two genes with two regulators. This  network is subject to different control modes of transcription, i.e., the genes are expressed by different mechanisms, repression, activation and a combination of them. The example is represented by the following algebraic system of non linear differential equations. In Figure 2, a schematic for the studied gene network is shown. $ X_{1} $ is an activator mRNA, $ X_{2} $ is the activator protein, $ X_{3} $ is the repressor mRNA, $ X_{4} $ is the repressor protein, $ X_{A} $ is the inmature activator and $ X_{R} $ is the immature repressor., see M. Savageau, \cite{LomnitzSavageau2015}.\\
\begin{figure}[htbp]
\centering
\makebox{\includegraphics[height=2.0in]{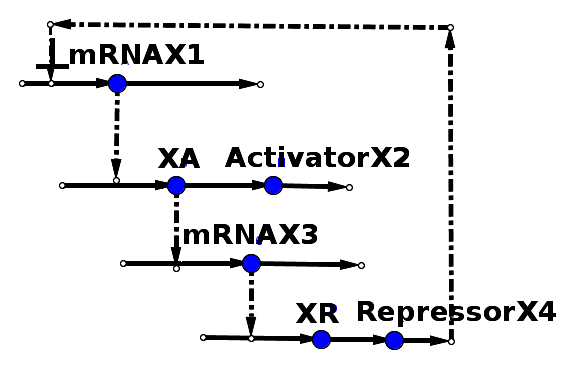}}
\caption{\label{fig_simvarex} Schematic representation of oscillatory gene circuit with six elements.}
\end{figure}

In Savageau and Lomnitz \cite{LomnitzSavageau2015}, the following variables are considered as follows: $X_{2i}=x_{2i}/(K^{0}_{2i}), \: i=1,2.$, and\\
\begin{center}
$ X_{2i-1}= \left(\dfrac{(\beta^{0}_{2i-1})(\sqrt{\rho^{0}_{2i-1}})}{\alpha^{0}_{2i-1}} \right)x_{2i-1}, \, i=1,2.$, 

\end{center}

where here the values of the concentrations of the immature activator and repressor were normalized with respect to its respectively values in steady-state, i.e, when we clear terms to solve equations in steady state, we have:\\

\begin{center}

$ X_{A}=\left[ \dfrac{(\beta^{0}_{1} + \mu^{0})(\beta^{0}_{A} +\mu^{0}) \sqrt{\rho^{0}_{1}}}{(\alpha^{0}_{1})(\alpha^{0}_{A})} \right] x_{A} $,

$ X_{R}=\left[\dfrac{(\beta^{0}_{3} + \mu^{0})(\beta^{0}_{R} + \mu^{0}) \sqrt{\rho^{0}_{3}}}{(\alpha^{0}_{3})(\alpha^{0}_{R})} \right] x_{R} $,
\end{center}
the regulators are normalized with respect to their dissociation constants of DNA. Near, their promoter regions, the same pair of binding sites for the repressor-DNA, and activator-DNA at both promoters are assumed, i.e, $ K_{2R}=K^{0}_{2} $ and $K_{4A}=K^{0}_{4} $, the parameters are explained below.\\

%$ \dot{X}_{1}= \left[  \alpha_{1} (\beta^{1}_{0} + \mu^{0}) \sqrt{\rho^{0}_{1}} (\alpha_{1}^{0})^{-1} \right] \lbrace  \rho_{1}^{- \pi_{1}} (X_{5})^{-1}$ 
%$+ \delta_{1}(K_{2})^{-g_{12}}(K^{0}_{2})^{g_{12}}X_{2}^{g_{12}} (X_{5})^{-1}  +$\\
% $(\rho_{1})^{-1} (K_{4A})^{-g_{14}} (K^{0}_{4})^{g_{14}}(X_{4})^{g_{14}} (X_{5})^{-1} \rbrace - (\beta_{1} + \mu) x_{1} $,\\\\
Now we give the explicit system of differential algebraic equations that represent the S-System associated to our gene-metabolic network , where this non linear system captures the advantages of essential non linear behaviour locally, the principal details about how to obtain this S-System can be consulted in Savageau and Lomintz \cite{LomnitzSavageau2015}. We here, only focus in analysing it by means of toric geometry.\\
This S-systems is a valid approximation of the full system if the conditions following
for the dominant terms are satisfied:\\\\

$ \dot{X}_{1}= \left[ \alpha_{1} (\beta^{1}_{0} + \mu^{0}) \sqrt{\rho^{0}_{1}} (\alpha_{1}^{0})^{-1} \right] \delta_{1}(K_{2})^{-g_{12}}(K^{0}_{2})^{g_{12}}X_{2}^{g_{12}} (K_{4A})^{g_{14}} (K_{4}^{0})^{-g_{14}} X_{4}^{-g_{14} } - (\beta_{1} + \mu) X_{1} $,\\\\
$ \dot{X}_{A}= (\beta^{0}_{A} + \mu^{0})\left[ \dfrac{\alpha_{A}}{\alpha^{0}_{A}} \right]  X_{1} - (\beta_{A} + \mu) X_{A}$,\\\\
$\dot{X}_{2} = (\beta^{0}_{2} + \mu^{0}) \left( \dfrac{\beta_{A}}{\beta^{0}_{A}}\dfrac{\alpha_{1}^{0} n_{2}^{0}}{(\beta_{1}^{0} + \mu^{0})\sqrt{\rho_{1}^{0}} K_{2}^{0}} \right) X_{A} - (\beta_{2} + \mu)X_{2} $,\\\\\\
%$ \dot{X}_{3}=\left[  \alpha_{3} (\beta^{3}_{0} + \mu^{0}) \sqrt{\rho^{0}_{3}} (\alpha_{3}^{0})^{-1} \right] \lbrace (\rho_{3})^{- \pi_{3}}(x_{6})^{-1} +
%(K_{2R})^{-g_{32}}(K_{2}^{0})^{g_{32}} (X_{2})^{g_{32}}(X_{6})^{-1} +
% \delta_{3}$\\
% $(\rho_{3})^{-1} (K_{4})^{-g_{34}}(K_{4}^{0})^{g_{34}}X_{4}^{g_{34}} (X_{6})^{-1}  \rbrace - (\beta_{3}+ \mu) X_{3} $,\\\\
$ \dot{X}_{3}= \left[  \alpha_{3} (\beta^{3}_{0} + \mu^{0}) \sqrt{\rho^{0}_{3}} (\alpha_{3}^{0})^{-1} \right]
(K_{2R})^{-g_{32}}(K_{2}^{0})^{g_{32}} (X_{2})^{g_{32}} - (\beta_{3}+ \mu) X_{3} $,\\\\
$ \dot{X}_{R}= (\beta^{0}_{R} + \mu^{0})\left[ \dfrac{\alpha_{R}}{\alpha^{0}_{R}} \right]  X_{1} - (\beta_{R} + \mu) X_{R}$,\\\\
$ \dot{X}_{4} = (\beta^{0}_{4} + \mu^{0}) \left( \dfrac{\beta_{R}}{\beta^{0}_{R}}\dfrac{\alpha_{3}^{0} n_{4}^{0}}{(\beta_{3}^{0} + \mu^{0})\sqrt{\rho_{3}^{0}} K_{4}^{0}} \right) X_{R} - (\beta_{4} + \mu)X_{4} $.\\\\\\
%$ 0= f_{1}= 1- X_{5}  + \delta_{1}(K_{2})^{-g_{12}}(K_{2}^{0})^{g_{12}}(X_{2})^{g_{12}}  +
 %(K_{4A})^{-g_{14}} (K_{4}^{0})^{g_{14}} X_{4}^{g_{14} }$\\\\
%$ 0=f_{2}= 1 - X_{6} +  (K_{2R})^{-g_{32}}(K_{2}^{0})^{g_{32}} (X_{2})^{g_{32}}%(X_{6})^{-1} + \delta_{3}(K_{4})^{-g_{34}}(K_{4}^{0})^{g_{34}}X_{4}^{g_{34}} $\\

where $ K_{2R} $ and $ K_{4A}$ are defined as:\\\\\\

$ K_{4A}= \dfrac{\left[  \dfrac{\alpha_{3}^{0} n_{4}^{0}}{(\beta_{3}^{0} + \mu^{0})\sqrt{\rho_{3}^{0}}} \right]}{\left\lbrace   \left( \delta_{1} \left[ \dfrac{\alpha_{1}^{0} n_{2}^{0}}{(\beta_{1}^{0} + \mu^{0})\sqrt{\rho_{1}^{0}} K_{2}^{0}} \right]^{g_{12}}  + 1 - \pi_{1}  \right) \sqrt{\rho_{1}^{0}} - \pi_{1} \right\rbrace^{\dfrac{1}{g_{14}}}} $;\\\\

$ K_{2R}=\dfrac{   \left[  \dfrac{ \alpha_{1}^{0} n_{2}^{0}  }{(\beta_{1}^{0}  + \mu^{0}) \sqrt{\rho_{1}^{0}} } \right] \sqrt{\rho_{3}^{0}}^{\dfrac{1}{g_{32}}}}{ \left\lbrace   \delta_{3} \left[  \dfrac{\alpha_{3}^{0} n_{4}^{0}}{(\beta_{3}^{0}  + \mu^{0})\sqrt{\rho_{3}^{0}} K_{4}^{0} }  \right]^{g_{34}}   + (\pi_{3} - 1)\sqrt{\rho_{3}^{0}} + \pi_{3} \right\rbrace^{\dfrac{1}{g_{32}}}   } $. \\\\\\
The details of this formulation are given with a wide development of the calculations, in; J. Lomintz and M. Savageau, \cite{LomnitzSavageau2015}.\\

The parameters space and all variables belong to the following Euclidian-space, $ X=(X_{1}, X_{2}, X_{3}, X_{4}, X_{A}, X_{R}) \, \in \, \mathbb{R}^{6}  $; X represents the concentration vector, and it contains the concentration for each of the elements of the system.\\

\begin{itemize}
\item $ X_{1}: $ Concentration of activator mRNA. 
\item $ X_{2}: $ Concentration of activator protein.
\item $ X_{3} :$ Concentration of repressor mRNA.
\item $ X_{4} :$ Concentration of repressor protein.
%\item $ X_{5} :$ Change of variable for the rational function of rate law associated to the kinetic enzyme of $ X_{1} $. 
%\item $ X_{6} :$ Change of variable for the rational function of rate law associated to the kinetic enzyme of $ X_{3} $. 
\item $ X_{A} :$ Concentration of form immature activator.
\item $ X_{R} :$ Concentration of form immature repressor.
\end{itemize} 
For this system, the condition lattice vector is defined by $ g=(g_{12}, g_{14}, g_{32}, g_{34}) \, \in \, \mathbb{Z}^{4} $, where the $ g_{ij}'s $, are kinetic parameters that represent the cooperativity of repression or activation and are associated to the number of binding sites occupied by the regulator protein.\\
The constants $ \alpha_{i}, \beta_{i}, \mu^{0}, \rho_{i}^{0}, n_{i}^{0} \, \in \, \mathbb{R}$ are: maximum rates of transcription, rate constants of mRNA and tagged proteins at midlevel expression, rates of dilution due to exponential growth, regulatory capacities, and ratio of proteins to mRNA, respectively. The constants $K_{2}, K_{4}, K_{2R}, K_{4A}  \, \in \, \mathbb{R}$, are concentrations of regulator for half-maximal binding to control sequences in the DNA, $ K_{2} $ and $ K_{4} $ for binding near their own promoter regions and $ K_{2R} $, $ K_{4A} $ for binding near other promotors, see, J. Lomintz and M. Savageau \cite{LomnitzSavageau2014}.\\
The parameters $ \pi_{i}, \, \delta_{i}= 1,0. \, \in \, \mathbb{Z}_{2}$, for $ i=1,2,3,4. $ represent the different modes of control for the activator and repressor proteins, For $ \pi_{i}= 1 $, the gene network is considered to behave as an activator-primary system, in which the default mode of the system is inactive, and the activator is the primary regulator needed to increase the expression from the promoter. On the other hand, in the case $ \pi_{i}= 0$, the behavior of the system is as a repressror-primary system, in which the promoter is fully active in the absence of both regulators and the repression is needed to reduce expression from the promoter. $  \delta_{i}=1$ will imply a double-regulator network and , and $ \delta_{i}=0 $ will imply a single-regulator network. The solution for the S-system, described previously, under different control modes is shown in the following figures.

%when $ \pi_{i}= 1 $, we will say that the gene network has a \textbf{primary-activator} in the case $ \pi_{i}= 0 $ it will be a \textbf{repressor-primary}. Now let us study the solution of the S-systems for different control modes represented by the following figures below, see; figs. 3-8:

\begin{center}
\textbf{Gene circuit with a single negative feedback loop.}\\
$ \delta_{1}=0, \delta_{3}=0, \pi_{1}=0, \pi_{3}=1 $.

\begin{figure}[htbp]
\centering
%\makebox{\includegraphics[height=1.75in]{C1.png}}
\includegraphics[height=1.60in]{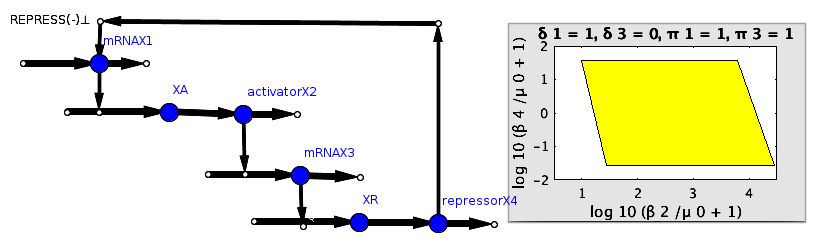}
\caption{\label{fig_simvarex} Single negative feedback loop.}
\end{figure}
\end{center}

\begin{center}
\textbf{Gene circuits with one positive and one negative feedback loop.}\\

$ \delta_{1}=1, \delta_{3}=0, \pi_{1}=1, \pi_{3}=1 $,

\begin{figure}[htbp]
\centering
\makebox{\includegraphics[height=1.60in]{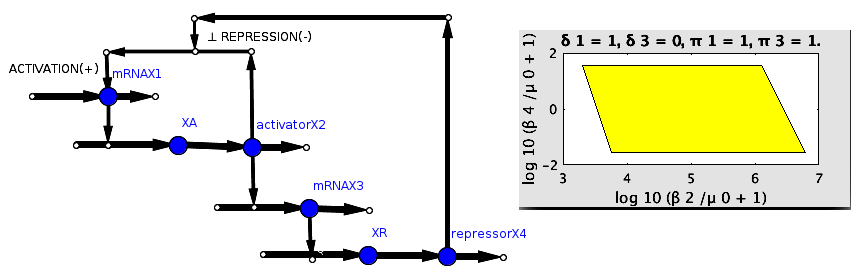}}

%\caption{\label{fig_simvarex} Positive feedback loop.}
\caption{\label{fig_simvarex} Activator-only control of repression transcription and activator-primary control of activator transcription.}

\end{figure}

\end{center}

\begin{center}
$ \delta_{1}=1, \delta_{3}=0, \pi_{1}=0, \pi_{3}=1 $.

\begin{figure}[htbp]
\centering
\makebox{\includegraphics[height=1.60in]{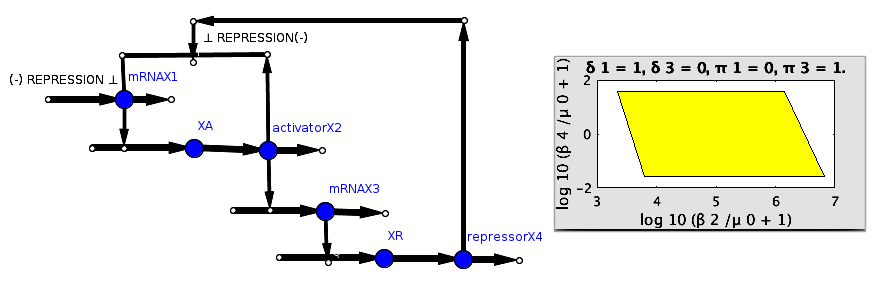}}
%\caption{\label{fig_simvarex} Negative feedback loops.}
\caption{\label{fig_simvarex} Activator-only control of repression transcription and repressor-primary control of activator transcription.}
\end{figure}

\end{center}

\begin{center}
\textbf{Gene circuits with one positive and one negative feedback loops.}\\

$ \delta_{1}=1, \delta_{3}=1, \pi_{1}=1, \pi_{3}=1 $,

\begin{figure}[htbp]
\centering
\makebox{\includegraphics[height=1.60in]{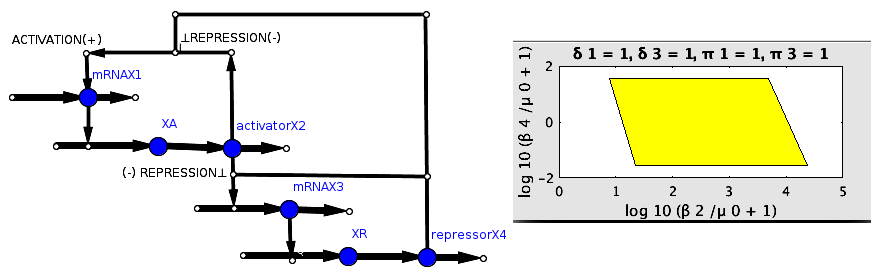}}
%\caption{\label{fig_simvarex} Positive loop: Activator-Primary control of repressor transcription.}
\caption{\label{fig_simvarex} Activator-primary control of repressor transcription and activator-primary control of activator transcription.}
\end{figure}

\end{center}

\begin{center}
$ \delta_{1}=1, \delta_{3}=1, \pi_{1}=0, \pi_{3}=1 $.

\begin{figure}[htbp]
\centering
\makebox{\includegraphics[height=1.60in]{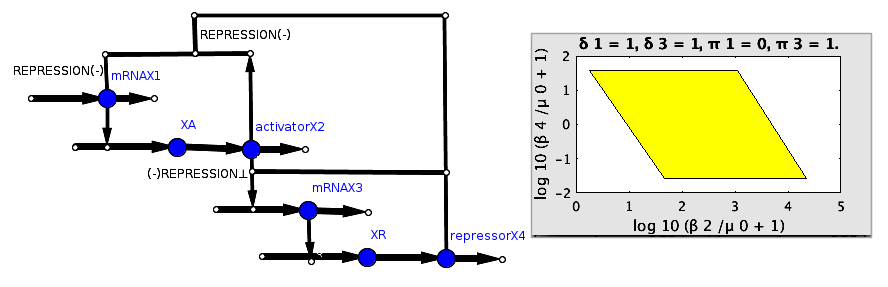}}
%\caption{\label{fig_simvarex} Nested negative feedback loop: Activator-Primary control of repressor transcription.}
\caption{\label{fig_simvarex} Activator-primary control of repressor transcription and repressor-primary control of activator transcription.}
\end{figure}

\end{center}

\begin{center}
\textbf{Gene circuits with one positive and nested negative feedback loops.}\\

$ \delta_{1}=1, \delta_{3}=1, \pi_{1}=1, \pi_{3}=0 $,

\begin{figure}[htbp]
\centering
\makebox{\includegraphics[height=1.60in]{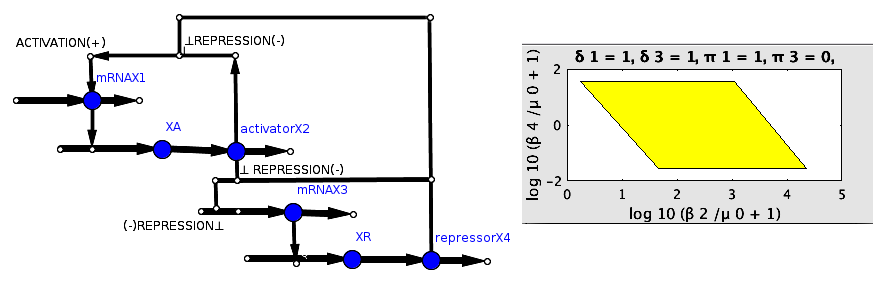}}
%\caption{\label{fig_simvarex} Positive loop: Repressor-Primary control of repressor transcription.}
\caption{\label{fig_simvarex} Repressor-primary control of repressor transcription and activator-primary control of activator transcription.}
\end{figure}

\end{center}

\begin{center}
$ \delta_{1}=1, \delta_{3}=1, \pi_{1}=0, \pi_{3}=0 $.

\begin{figure}[htbp]
\centering
\makebox{\includegraphics[height=1.60in]{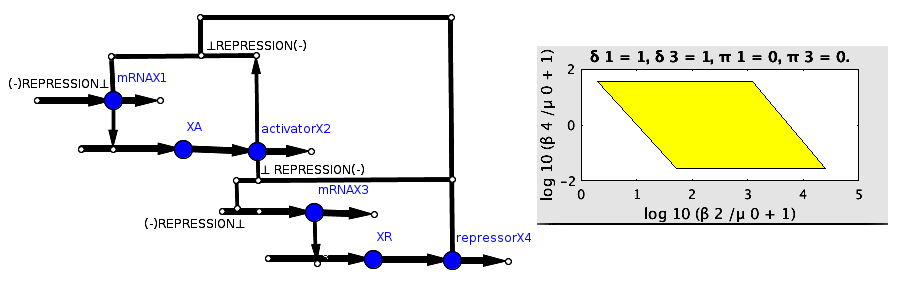}}
%\caption{\label{fig_simvarex} Nested negative feedback loops: Repressor-Primary control of repressor transcription.}
\caption{\label{fig_simvarex} Repressor-primary control of repressor transcription and repressor-primary control of activator transcription.}
\end{figure}

\end{center}

%\textbf{Applications of Computational Toric S-Systems: Phenotypic toric ideals and phenotypic variety. }\\\\\\

%\subsection{Applications of computational Toric Geometry}\
%\label{sec:figures}
Each one of the different control modes (represented as a combination of $ \delta's $ and $ \pi's $) is mapped to complex geometric regions, which are coloured in yellow in Figures 3-9. In a study done by Lomnitz and Savageau \cite{LomnitzSavageau2014}, these regions correspond to one complex solution associated to an eigenvalue with real positive part. In Savageau's work, a specific region of the possible solution-space of this system corresponds to an oscillatory region. In our analysis, performed using Hilbert basis and toric ideals, we partially reproduce these results and we found that part of the oscillatory region is conserved across most of the different control modes, this conserved region is called a conserved \textbf{phenotypic polytope}. This technique is important since it would allow the modelling of gene circuits with hundreds of interacting nodes, with the possibility to find \textbf{conserved phenotype polytopes} within such regulatory networks.\\

%%discusion va
%is not our aim interest to solve the S-systems in this work, however, in a second article will do it by means of Theorem 8 (toric resolution in S-systems), which is a technique different to classical methods. 

\textbf{Methodological analysis of S-systems by means of toric varieties.}

%The S-systems defined previously will be analysed in the last part of this work, and we start considering these equations in \textbf{steady-state} that is to say. When $  \dot{X}_{2} =\dot{X}_{4}=0 $, since that we are only interesting in the study of control of primary activator $ X_{2} $ , for the transcription of repressor $ X_{4}$. 
We will start by considering an S-System in steady-state. In such state there is no change in the concentration for every species of the network, i.e. $  \dot{X}_{2} =\dot{X}_{4}=0 $. We are only interested in$ X_{2} $ and $ X_{4} $ since $ X_{2} $ is the activator protein and $ X_{4} $ is the repressor protein.
This set of polynomials is an $ \mathbb{C}-$\textbf{algebra} well defined, $ \mathbb{C}[X_{i}^{g_{ij}}, X_{i}^{h_{ij}}] $; by means of theorem 3.6\label{Theorem 2.3.0.}, the toric ideals are finitely generated and have also an associated geometric realization of lattice polytope as we will see. Our coordinate ring is defined in the variables $ X=(X_{1}, X_{2}, X_{3}, X_{4}, X_{A}, X_{R}) \, \in \, \mathbb{R}^{6}$ and is oriented with respect to a canonical basis $ B=\lbrace e_{1}, e_{2}, e_{3}, e_{4}, e_{5}, e_{6} \rbrace \, \subseteq \, \mathbb{R}^{6}$. To construct the Newton polytope from the S-sytems, the next methodological development is used: \\
1). \textbf{Computing of the support-space for each equation of the S-System}, we have agreement of the definition of Newton-polyhedron or exponent-space associated to each equations to S-sytems, the following sets can be therefore stipulated:\\
%$ supp(\dot{X}_{1}) =\left\lbrace  -e_{5}, g_{12}e_{2} -e_{5}, g_{14}e_{4}-e_{5}, e_{1}   \right\rbrace $.\\
$ supp(\dot{X}_{1}) =\left\lbrace g_{12}e_{2}-g_{14}e_{4}, e_{1} \right\rbrace $.\\
$ supp(\dot{X}_{A}) =\left\lbrace e_{1}, e_{5} \right\rbrace $.\\
$ supp(\dot{X}_{2}) =\left\lbrace e_{5}, e_{2} \right\rbrace $.\\
$ supp(\dot{X}_{3}) =\left\lbrace g_{32}e_{2}, e_{3}\right\rbrace $.\\
$ supp(\dot{X}_{R}) =\left\lbrace e_{3}, e_{6}\right\rbrace $.\\
$ supp(\dot{X}_{4}) =\left\lbrace e_{6}, e_{4}\right\rbrace $.\\\\

%$ supp(\dot{X}_{5}) =\left\lbrace g_{12} e_{2}, g_{14} e_{4}, e_{5}    \right\rbrace $.\\
%$ supp(\dot{X}_{6}) =\left\lbrace g_{32} e_{2},g_{34} e_{4}, e_{6}    \right\rbrace $.\\
2.) \textbf{Computing of the Hilbert basis for the monoids associated to the dual lattice cones.} Remember that the basis were obtained with the Singular program. Is important to note that the Hilbert basis are invariant for all multiplicative constant, for more details of this argument, ref. \cite{DGPS}, for example, for the lattice vector ray $e_{2} \, \in \, \mathbb{R}^{6} $ belongs to the elongated lattice vector ray $ g_{12}e_{2} \, \in \, \mathbb{R}^{6}$ for some constant $ g_{12} \, \in \, \mathbb{R} $, as is the case of the parameter values $ g_{12}, g_{14}, g_{32} $ and without more problem we adopt the values $g_{12}=g_{14}=g_{32} = 1 $, and we have.\\

Monoid $S_{\sigma^{\vee}_{1}}= \sigma^{\vee}_{1} \cap \mathbb{Z}^{6} $, Hilbert basis associated, $ H_{\sigma^{\vee}_{1}} =\left\lbrace  e_{1}, e_{2}- e_{4}\right\rbrace $.\\

Monoid $S_{\sigma^{\vee}_{A}}= \sigma^{\vee}_{1} \cap \mathbb{Z}^{6} $, Hilbert basis associated, $ H_{\sigma^{\vee}_{A}} =\left\lbrace  e_{1}, e_{5}  \right\rbrace $.\\

Monoid $S_{\sigma^{\vee}_{2}}= \sigma^{\vee}_{2} \cap \mathbb{Z}^{6} $, Hilbert basis associated, $ H_{\sigma^{\vee}_{2}} =\left\lbrace e_{5}, e_{2} \right\rbrace $.\\

Monoid $S_{\sigma^{\vee}_{3}}= \sigma^{\vee}_{3} \cap \mathbb{Z}^{6} $, Hilbert basis associated, $ H_{\sigma^{\vee}_{3}} =\left\lbrace  e_{2}, e_{3} \right\rbrace $.\\

Monoid $S_{\sigma^{\vee}_{R}}= \sigma^{\vee}_{R} \cap \mathbb{Z}^{6} $, Hilbert basis associated, $ H_{\sigma^{\vee}_{R}} =\left\lbrace  e_{3}, e_{6}  \right\rbrace $.\\

Monoid $S_{\sigma^{\vee}_{4}}= \sigma^{\vee}_{4} \cap \mathbb{Z}^{6} $, Hilbert basis associated, $ H_{\sigma^{\vee}_{4}} =\left\lbrace e_{6}, e_{4}  \right\rbrace $.\\

%Monoid $S_{\sigma^{\vee}_{5}}= \sigma^{\vee}_{5} \cap \mathbb{Z}^{8} $, basis Hilbert associated, $ H_{\sigma^{\vee}_{5}} =\left\lbrace e_{1}, e_{6}, e_{3}, e_{7}  ,e_{8}   \right\rbrace $.\\

%Monoid $S_{\sigma^{\vee}_{6}}= \sigma^{\vee}_{6} \cap \mathbb{Z}^{8} $, Hilbert basis associated, $ H_{\sigma^{\vee}_{6}} =\left\lbrace  e_{1}, e_{3}, e_{7}, e_{8}  ,e_{5}  \right\rbrace $.\\

3). \textbf{Geometrical realizations of Newton polytopes and dual cones}. We write the notation of dualCone but we don't explicitly compute the lattice vectors that generate the dual cones, these are computed implicitly in the Singular program \cite{DGPS}. \\

$\sigma^{\vee}_{1}=$ dualCone $(g_{12}e_{2}-g_{14}e_{4}, e_{1}) $,\\
$ \sigma^{\vee}_{A}=$ dualCone $(e_{1}, e_{5} ) $,\\
$ \sigma^{\vee}_{2}=$ dualCone $(e_{5}, e_{2})$,\\
$ \sigma^{\vee}_{3}=$ dualCone $( g_{32}e_{2}, e_{3})$,\\
$ \sigma^{\vee}_{R}=$ dualCone $( e_{3}, e_{6}) $,\\
$ \sigma^{\vee}_{4}=$ dualCone $( e_{6}, e_{4})$.\\\\

%$ \sigma^{\vee}_{5}=$dualCone $(g_{12} e_{2}, g_{14} e_{4}, e_{5}) $,\\
%$ \sigma^{\vee}_{6}=$dualCone $(g_{32} e_{2},g_{34} e_{4}, e_{6}) $. \\
Let us remember that the geometric realization of a toric variety is made by dual cones since in this space the cones have the same dimension, and the gluing maps are well defined, see, ref. \cite{GEwald1993}. The computing of the dual cones is easy to see from the Hilbert basis for few elements or from these toric ideals. Therefore, the dual cones are:\\\\
$\sigma^{\vee}_{1}=$ Con  $(e_{3}, e_{5}, e_{6}, e_{1}+e_{4}, e_{1}+e_{2}) $,\\
$\sigma^{\vee}_{A}=$ Con $(e_{2}, e_{3}, e_{4}, {e}_{6}, e_{1}+e_{5})$,\\
$\sigma^{\vee}_{2}=$ Con $(e_{1}, e_{3}, e_{4}, e_{6},e_{2}+e_{5})$,\\
$\sigma^{\vee}_{3}=$ Con $(e_{1}, e_{4}, e_{5}, e_{6}, e_{2}+e_{3} )$,\\
$\sigma^{\vee}_{R}=$ Con $(e_{1}, e_{2}, e_{4}, e_{5}, e_{2}+e_{6} ) $,\\
$\sigma^{\vee}_{4}=$ Con $(e_{1}, e_{2}, e_{3}, e_{5}, e_{3}+e_{6} )$.\\
These were calculated from the toric ideals, once computed in 5).\\

4.) \textbf{Computing of toric ideals associated to the Hilbert basis.} 
The toric ideals for $ \langle \dot{X}_{A} \rangle,\: \langle \dot{X}_{R} \rangle,\: \langle \dot{X}_{1} \rangle,\: \langle \dot{X}_{2} \rangle, \: \langle \dot{X}_{3} \rangle,\: \langle \dot{X}_{4} \rangle$,  In these equations it is easy to calculate these values for some parameters $ h_{i,j}=1 $ for all $ i, j=1, 2, 3, 4, A, R $. Explicitly, the parameters can take the  values, for the same rationale applied in 2), $ g_{ij}= \lbrace g_{12}= g_{14}=g_{A1}=1, g_{2A}=1, g_{32}=g_{34}= g_{R3}=1, g_{4R}=1 \rbrace $ generating all the possible connections for different control modes in this gene circuit. And were computed by means of the technique of \textbf{Hilbert basis} in agreement with the theorems previously established (Theorem 4.4\label{Theorem 1.} and Theorem 4.5\label{Theorem 2.} (toric resolution in S-systems)). Their computing was made with the program, \textbf{Singular} \cite{DGPS}. Before it is necessary to renormalise the variables of S-system as input for the Singular program, hence:\\
We rename the parameters for simplicity and after rename the variables abusing of notation, being that the toric ideals associated to each equation are the same if the variable is multiplied for different parameter.\\
$\dot{X}_{1} =a*X_{2}X_{4}^{-1}- b*X_{1}  $,\\
$\dot{X}_{A} =c*X_{1}- d*X_{A} $,\\
$\dot{X}_{2} =e*X_{A}- f*X_{2} $,\\
$\dot{X}_{3} =g*X_{2}- h*X_{3} $,\\
$\dot{X}_{R} =i*X_{3}- j*X_{R} $,\\
$\dot{X}_{4} =k*X_{R}- l*X_{4} $.\\

And rename variables, also let us to do abuse of the notation here, for each variable $ u_{i} $, as we see.\\
$\dot{u}_{1} =u_{2}u_{4}^{-1}- u_{1}  $,\\
$\dot{u}_{A} =u_{1}- u_{A} $,\\
$\dot{u}_{2} =u_{A}- u_{2} $,\\
$\dot{u}_{3} =u_{2}- u_{3} $,\\
$\dot{u}_{R} =u_{3}- u_{R} $,\\
$\dot{u}_{4} =u_{R}- u_{4} $.\\
%Where the derivaties are defined as $\dot{u}_{i}= C_{i}\dot{X}_{i}  $ with $ C_{i}'s \, \in \, \mathbb{R}$ real constants after of the normalisation, and we abuse of the notation, an 
We do simply $\dot{u}_{i}= \dot{X}_{i}  $, being that finally will do $ \dot{X}_{i}=0 $ (steady-state condition), so the toric ideals are:\\ 
%$\langle \dot{X}_{1} \rangle= \langle X_{A}-1, X_{2}*X_{4}-1, X_{3}- 1, X_{R}-1 \rangle $,\\
$I_{\sigma^{\vee}_{1}}=\langle \dot{X}_{1} \rangle= \langle u_{1}u_{4}-1, u_{1}u_{2}-1, u_{3}- 1, u_{A}-1, u_{R}-1 \rangle $,\\
%$\langle \dot{X}_{A} \rangle= \langle X_{R}-1, X_{2}-1, X_{3}- 1, X_{4}- 1\rangle  $,\\
$I_{\sigma^{\vee}_{A}}=\langle \dot{X}_{A} \rangle= \langle u_{1}u_{A}-1, u_{R}-1, u_{2}-1, u_{3}- 1, u_{4}- 1\rangle  $,\\
%$\langle \dot{X}_{2} \rangle= \langle X_{1}-1, X_{R}-1, X_{3}- 1, X_{4}- 1 \rangle  $,\\
$I_{\sigma^{\vee}_{2}}=\langle \dot{X}_{2} \rangle= \langle u_{2}u_{A}-1, u_{1}-1, u_{R}-1, u_{3}- 1, u_{4}- 1 \rangle  $,\\
%$\langle \dot{X}_{3} \rangle= \langle X_{1}-1, X_{R}- 1, X_{A}- 1, X_{4}- 1 \rangle  $,\\
$I_{\sigma^{\vee}_{3}}=\langle \dot{X}_{3} \rangle= \langle u_{2}u_{3}-1, u_{1}-1, u_{R}- 1, u_{A}- 1, u_{4}- 1 \rangle  $,\\
%$\langle \dot{X}_{R} \rangle= \langle X_{1}-1, X_{2}-1, X_{A}- 1, X_{4}- 1 \rangle  $,\\
$I_{\sigma^{\vee}_{R}}=\langle \dot{X}_{R} \rangle= \langle u_{2}u_{R}-1, u_{1}-1, u_{2}-1, u_{A}- 1, u_{4}- 1 \rangle  $,\\
%$\langle \dot{X}_{4} \rangle= \langle X_{1}-1, X_{2}-1, X_{3}- 1, X_{A}- 1 \rangle  $,\\
$I_{\sigma^{\vee}_{4}}=\langle \dot{X}_{4} \rangle= \langle u_{4}u_{R}-1, u_{1}-1, u_{2}-1, u_{3}- 1, u_{A}- 1 \rangle  $.\\
%$\langle {f}_{1} \rangle= \langle X_{1}-1, X_{6}-1, X_{3}- 1, X_{A}- 1, X_{R}- 1 \rangle  $,\\
%$\langle {f}_{2} \rangle= \langle X_{1}-1, X_{3}-1, X_{R}- 1, X_{A}- 1, X_{5} -1 \rangle  $.\\\\

Therefore the fan $ \sum=\lbrace \sigma^{\vee}_{i} : \, i=1,2,3,4,A,R \rbrace $ built with this system of lattice cones, yields the phenotypic variety given by:
\begin{center}
$  X_{\sigma^{\vee}_{i}} =$\textbf{Spec} $( \mathbb{C}[X_{j}^{g_{ij}}, X_{j}^{h_{ij}}])$, $ \: \forall \, i, j=1,2,3,4, A, R. $
\end{center}

5.) \textbf{Realization of the toric varieties associated to the dual cones.} The spectrum or set of prime ideals associated to the toric varieties, is our new coordinate system for translating and solving the S-system associated to any gene-metabolic network. In particular, we are interested in the gene circuit described earlier for reproducing the oscillatory phenotype associated to them, and given in Savageau \cite{LomnitzSavageau2014}.\\

More explicitly the toric varieties are:\\\
%$X_{\sigma^{\vee}_{1}} =$\textbf{Spec} $( \mathbb{C}[\sigma^{\vee}_{1} \cap \mathbb{Z}^{6}])= \left\lbrace X_{A}, X_{R}, X_{3}, X_{6} \right\rbrace $,\\
%$  X_{\sigma^{\vee}_{2}} = $\textbf{Spec} $( \mathbb{C}[\sigma^{\vee}_{2} \cap \mathbb{Z}^{6}])=\left\lbrace X_{1}, X_{R}, X_{3}, X_{4}, X_{5}, X_{6} \right\rbrace$,\\
$X_{\sigma^{\vee}_{1}} =$\textbf{Spec} $( \mathbb{C}[\sigma^{\vee}_{1} \cap \mathbb{Z}^{6}])= 
$\textbf{Spec} $( \mathbb{C}[u_{1},u_{2}u^{-1}_{4}]) \\
= 
\left\lbrace u_{1},u_{2}u^{-1}_{4} \right\rbrace \simeq \mathbb{C}_{(u_{1})}\times \mathbb{C}_{(u_{2})} \times \mathbb{CP}_{(u_{4})} $,\\
$  X_{\sigma^{\vee}_{2}} =$\textbf{Spec} $( \mathbb{C}[\sigma^{\vee}_{2} \cap \mathbb{Z}^{6}])= $\textbf{Spec} $( \mathbb{C}[u_{2},u_{A}])\\
=\left\lbrace u_{2},u_{A} \right\rbrace \simeq \mathbb{C}^{2}_{(u_{2},u_{A})}$,\\
$  X_{\sigma^{\vee}_{3}} = $\textbf{Spec} $( \mathbb{C}[\sigma^{\vee}_{3} \cap \mathbb{Z}^{6}])=$\textbf{Spec} $( \mathbb{C}[u_{2},u_{3}])\\
= \left\lbrace u_{2},u_{3}\right\rbrace \simeq \mathbb{C}^{2}_{(u_{2},u_{3})}$,\\
$  X_{\sigma^{\vee}_{4}} =$\textbf{Spec} $( \mathbb{C}[\sigma^{\vee}_{4} \cap \mathbb{Z}^{6}])=$\textbf{Spec} $( \mathbb{C}[u_{4},u_{R}])\\
=\left\lbrace u_{4}, u_{R} \right\rbrace \simeq \mathbb{C}^{2}_{(u_{4},u_{R})}$,\\
%$  X_{\sigma^{\vee}_{5}} =$\textbf{Spec} $( \mathbb{C}[\sigma^{\vee}_{5} \cap \mathbb{Z}^{8}])=\left\lbrace X_{1}, X_{3}, X_{6}, X_{A}, X_{R} \right\rbrace$,\\
%$  X_{\sigma^{\vee}_{6}} =$\textbf{Spec} $( \mathbb{C}[\sigma^{\vee}_{6} \cap \mathbb{Z}^{8}])=\left\lbrace X_{1}, X_{3}, X_{5}, X_{A}, X_{R} \right\rbrace$,\\
$  X_{\sigma^{\vee}_{A}} =$\textbf{Spec} $( \mathbb{C}[\sigma^{\vee}_{A} \cap \mathbb{Z}^{6}])=$\textbf{Spec} $( \mathbb{C}[u_{1}, u_{A}])\\
=\left\lbrace u_{1}, u_{A} \right\rbrace\simeq \mathbb{C}^{2}_{(u_{1},u_{A})}$,\\
$  X_{\sigma^{\vee}_{R}} =$\textbf{Spec} $( \mathbb{C}[\sigma^{\vee}_{R} \cap \mathbb{Z}^{6}])=$\textbf{Spec} $( \mathbb{C}[u_{3},u_{R}])\\
=\left\lbrace u_{3}, u_{R} \right\rbrace \simeq \mathbb{C}^{2}_{(u_{3},u_{R})}$.\\\\
\textbf{Gluing toric varieties.} Now we glue these toric varieties through of the gluing maps defined by means of their common faces in each one of the dual cones\\
\begin{center}
$ \psi_{\sigma_{1},\sigma_{A}}= X_{\sigma^{\vee}_{1}} / \lbrace u_{3}=u_{R}= 0\rbrace \longrightarrow X_{\tau_{1}}\longrightarrow   X_{\sigma^{\vee}_{A}} / \lbrace u_{3}=u_{R}= 0\rbrace $,\\

$ \psi_{\sigma_{A},\sigma_{2}}= X_{\sigma^{\vee}_{A}} / \lbrace u_{3}=u_{4}=u_{R}= 0\rbrace \longrightarrow X_{\tau_{A}}\longrightarrow   X_{\sigma^{\vee}_{2}} / \lbrace u_{3}=u_{4}=u_{R}= 0\rbrace $,\\

$ \psi_{\sigma_{2},\sigma_{3}}= X_{\sigma^{\vee}_{2}} / \lbrace u_{1}=u_{4}=u_{6}= 0\rbrace \longrightarrow X_{\tau_{2}}\longrightarrow   X_{\sigma^{\vee}_{3}} / \lbrace u_{1}=u_{4}=u_{6}= 0\rbrace $,\\

$ \psi_{\sigma_{3},\sigma_{R}}= X_{\sigma^{\vee}_{3}} / \lbrace u_{1}=u_{4}=u_{5}= 0\rbrace \longrightarrow X_{\tau_{3}}\longrightarrow   X_{\sigma^{\vee}_{R}} / \lbrace u_{1}=u_{4}=u_{5}= 0\rbrace $,\\
$ \psi_{\sigma_{R},\sigma_{4}}= X_{\sigma^{\vee}_{R}} / \lbrace u_{1}=u_{2}=u_{5}=0\rbrace \longrightarrow X_{\tau_{4}}\longrightarrow   X_{\sigma^{\vee}_{4}} / \lbrace u_{1}=u_{2}=u_{5}=0 \rbrace $,\\
$ \psi_{\sigma_{4},\sigma_{A}}= X_{\sigma^{\vee}_{4}} / \lbrace u_{3}=u_{5}= 0\rbrace \longrightarrow X_{\tau_{R}}\longrightarrow   X_{\sigma^{\vee}_{A}} / \lbrace u_{3}=u_{5}= 0\rbrace $.\\
\end{center}
Where the sub-varieties $ X_{\tau^{\vee}_{1}}$, $ X_{\tau^{\vee}_{A}}$, $ X_{\tau^{\vee}_{2}}$, $ X_{\tau^{\vee}_{3}}$, $ X_{\tau^{\vee}_{R}} $, $ X_{\tau^{\vee}_{4}}$ are given by:\\
%\begin{center}
$ X_{\tau_{1}}=X_{\sigma^{\vee}_{1}}/\lbrace z^{e_{1}}=u_{1}=0\rbrace \simeq \mathbb{C}^{2}_{(u_{3},u_{R})}\times \mathbb{C}_{(u_{1})}$,\\
$ X_{\tau_{A}}=X_{\sigma^{\vee}_{A}}/\lbrace z^{e_{5}}=u_{5}=0 \rbrace \simeq \mathbb{C}^{3}_{(u_{3},u_{4},u_{R})}\times \mathbb{C}_{(u_{5})}$,\\
$ X_{\tau_{2}}=X_{\sigma^{\vee}_{2}}/ \lbrace z^{e_{2}}=u_{2}=0 \rbrace \simeq \mathbb{C}^{3}_{(u_{1},u_{4},u_{6})}\times \mathbb{C}_{(u_{2})}$,\\
$X_{\tau_{3}}= X_{\sigma^{\vee}_{3}}/ \lbrace z^{e_{3}}=u_{3}=0 \rbrace \simeq \mathbb{C}^{3}_{(u_{1},u_{4},u_{5})}\times \mathbb{C}_{(u_{3})}$,\\
$ X_{\tau_{R}}=X_{\sigma^{\vee}_{R}}/ \lbrace z^{e_{6}}=u_{6}=0 \rbrace \simeq \mathbb{C}^{3}_{(u_{1},u_{2},u_{5})}\times \mathbb{C}_{(u_{6})}$,\\
$ X_{\tau_{4}}=X_{\sigma^{\vee}_{4}}/ \lbrace z^{-e_{4}}=u^{-1}_{4}=0 \rbrace \simeq \mathbb{C}^{2}_{(u_{3},u_{5})}\times \mathbb{C}^{*}_{(u^{-1}_{4})}$.
%\end{center}

With common faces are given by the intersections in the generators of each monoid $S_{\sigma^{\vee}_{i}}=\sigma^{\vee}_{1}\cap \mathbb{Z}^{6}  $ with $ i=1,2,3,4,A;R $, i.e., common elements in the Hilbert basis, hence, $ \tau_{1}=\lbrace e_{1}\rbrace,\tau_{A}=\lbrace e_{5} \rbrace ,\tau_{2}=\lbrace e_{2} \rbrace ,\tau_{3}=\lbrace e_{3} \rbrace  ,\tau_{R}=\lbrace e_{6} \rbrace ,\tau_{4}=\lbrace e_{4}\rbrace $.\\

\begin{figure}[htbp]
\centering
\makebox{\includegraphics[height=2.2in]{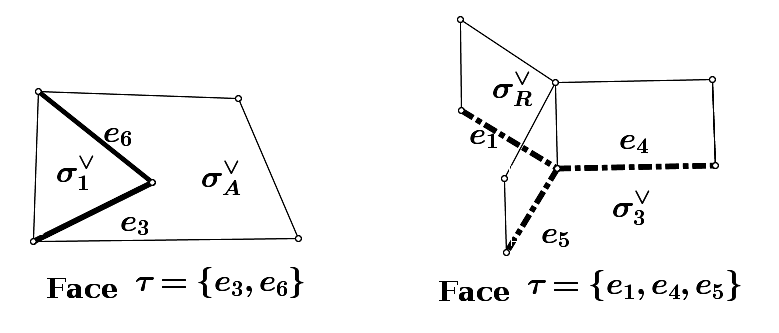}}
\caption{\label{fig_simvarex} The dual cones $ \sigma^{\vee}_{1} $ and $ \sigma^{\vee}_{A} $ are gluing through the face generated by $ e_{3},e_{6} $, similar for the cones $ \sigma^{\vee}_{3} $ and $ \sigma^{\vee}_{R} $, the cone $ \sigma^{\vee}_{R} $ intersect for three hyperplanes to the dual cone $ \sigma^{\vee}_{3} $ and the edges $ e_{1},e_{4},e_{5} $ are common faces to them.}
\end{figure}

%$ \sigma^{\vee}_{5}=$ Con $(e_{1}, e_{6}, e_{3}, e_{7}, e_{8}) $,\\
%$ \sigma^{\vee}_{6}=$ Con $(e_{1}, e_{3}, e_{5}, e_{7}, e_{8}) $.\\\\

The \textbf{phenotypic variety} is obtained with these gluing maps, and the fan defined as $ \sum= \cup_{i} \sigma^{\vee}_{i}$, $ i=1,2,3,4,A,R. $. Then we have the following toric variety.
\begin{center}
$  X_{\sum} =X_{\sigma^{\vee}_{1}}\cup X_{\sigma^{\vee}_{2}}\cup X_{\sigma^{\vee}_{3}} \cup X_{\sigma^{\vee}_{4}} \cup X_{\sigma^{\vee}_{A}} \cup X_{\sigma^{\vee}_{R}}$.
\end{center}
Also is true the realization of general toric variety associated to the S-system in this case:
$ X_{\sum}= $\textbf{Spec} $( \mathbb{C}[\sum \cap \mathbb{Z}^{6}])=$\textbf{Spec}$( \mathbb{C}[u_{1},u_{2},u_{3},u_{4},u_{A},u_{R}]/ I_{\sum})\\
=\left\lbrace u_{2}=u^{2}_{4} \right\rbrace$.

Substituting all the algebraic relations, we obtain the parabolic relationship between immature regulators, $ u_{2}=u^{2}_{4} $, in the original variables:\\

$ a*X_{2}=bc^{-1}de^{-1}fg^{-1}i^{-1}jk^{-1}l*X^{2}_{4} $, or only $X_{2}=\tilde{K}*b*l* X^{2}_{4} $;\\

where $ b=(\beta_{2} + 1) $ and $l= (\beta_{4} + 1) $ important parameters considered in the next section, with $\tilde{K}  $ depending the other parameters and control modes $ \pi_{i}, \, \delta_{i} $. The normalised concentrations $ u_{2} $ and $ u_{4} $ always satisfies the condition in steady-state, this affine algebraic set, can be calculated easily simultaneously solving the parametric system equal to zero, i.e., $ \dot{X}_{1}=...=\dot{X}_{A}=\dot{X}_{R}=0 $, since the equivalence classes satisfied the equality.\\
In this point is important to mention, that the oscillatory genetic metabolic network can be completely described it by means of gluing maps which are toric morphisms of the dual cones into regular fan $ \sum= \cup_{i} \sigma^{\vee}_{i},\,i=1,2,3,4,A,R$. They are associated to monomial parametrizations of their respective differential equation in the S-system, for this genetic metabolic network  \textbf{(GMN)}. 
\begin{center}
$  GMN=\left\lbrace \psi_{\sigma_{1},\sigma_{A}}, \psi_{\sigma_{A},\sigma_{2}}, \psi_{\sigma_{2},\sigma_{3}}, \psi_{\sigma_{3},\sigma_{R}},\psi_{\sigma_{R},\sigma_{4}}, \psi_{\sigma_{4},\sigma_{1}} \,\vert \, \sum =\cup_{i} \sigma^{\vee}_{i} , regular fan\right\rbrace  $ .
\end{center}
We claim that any genetic metabolic network described by this mathematical modelling (S-systems) can be represented by gluing maps or toric morphisms, and we write.\\

\begin{center}
$  GMN=\left\lbrace  \psi_{\sigma_{i},\sigma_{j}},  \, \vert \,\sigma_{i},\sigma_{j}\,\in\, \sum=\cup_{k=1}^{n} \sigma^{\vee}_{k} \right\rbrace $.
\end{center}

with $ \sum \, \subseteq \,\mathbb{R}^{n} $ regular fan, if the S-system consist of an irregular fan, then we apply the Hilbert basis technique. \\
Also following the central dogma of the molecular biology (which states that information only flows from DNA to RNA to protein), any genetic-metabolic system can be represented in a cascade network of fluxes of biomass (as shown above). This cascade of information is well represented geometrically and algebraically for the concatenation of gluing maps, as is shown below:
\begin{center}
$  \psi_{\sigma_{1},\sigma_{A}}\longrightarrow \psi_{\sigma_{A},\sigma_{2}}\longrightarrow \psi_{\sigma_{2},\sigma_{3}}\longrightarrow \psi_{\sigma_{3},\sigma_{R}} \longrightarrow \psi_{\sigma_{R},\sigma_{4}} \longrightarrow \psi_{\sigma_{4},\sigma_{1}}  $.
\end{center}
Each one of these maps contains algebraic information for re-parametrization of the binomials in the S-system, as we showed in the construction of the toric varieties.\\

\textbf{Obtaining a sustained oscillatory phenotypic region.}\\

In this section, we will obtain the complex region, studied by M. Savageau \cite{LomnitzSavageau2014}, which represents a sustained oscillatory phenotype. This phenotypic region is preserved under different control modes of the transcriptional regulation for this gene network. 
The results presented in this work only include the steady-state solution of the system. Even though only we work with the dynamical solution of the regulator gene $ X_{R} $ which is suggested to intersect the toric ideals with algebraic set most common at all, in this case were the ideals $ <u_{A} - 1> $ and $ <u_{R} - 1> $, as is shown below. 

%But is important to mention that the results here reproduced only include the steady-state, and therefore we take the dynamical solutions reported in the work developed by Savageau, for the change of concentrations in the activator and regulator gene $ X_{R} $ and $ X_{A} $, and the oscillating is reproduced for the case these solution are complex eigenvalues with positive real-part, as we will see. \\

We are interested in obtaining common phenotypic polytopes (phenotypic polytope). Intersecting the toric ideals gives the ideal most common to the complete gene circuit provided this intersection is not empty. A geometric interpretation is that we want common faces into the lattice polytope, these common faces are hyperplanes of concentrations that are sustained or conserved during all the steady-state available to the gene network. Also is important to say that this phenotypic polytope is the sustained oscillatory phenotype reproduced for the case where the solution has complex eigenvalues with positive real-part, shown by Savegau \cite{LomnitzSavageau2014}, and now we reproduce the result with toric ideals, the toric ideal associated to this oscillatory region is named as a \textbf{phenotypic toric ideal}.

Then the intersection of the ideals with Zariski closure not empty, yields:
\begin{center}
$ I_{P_{1}}=\langle \dot{X}_{A} \rangle \cap \langle \dot{X}_{1} \rangle \cap \langle \dot{X}_{2} \rangle \cap \langle \dot{X}_{3} \rangle = I_{P_{1}}=\langle {u}_{R} - 1 \rangle $,\\
$ I_{P_{2}}=\langle \dot{X}_{R} \rangle \cap \langle \dot{X}_{1} \rangle \cap \langle \dot{X}_{3} \rangle \cap \langle \dot{X}_{4} \rangle = I_{P_{2}}=\langle {u}_{A} - 1 \rangle $.
\end{center}
or in a geometric manner,
\begin{center}
$  \sigma^{\vee}_{P_{1}}=\sigma^{\vee}_{A} \cap \sigma^{\vee}_{2} \cap \sigma^{\vee}_{3} \cap \sigma^{\vee}_{R} \cap \sigma^{\vee}_{4}= $ Con $(e_{5})$,\\
$  \sigma^{\vee}_{P2}=\sigma^{\vee}_{R} \cap \sigma^{\vee}_{1} \cap \sigma^{\vee}_{3}  \cap \sigma^{\vee}_{4}= $ Con $(e_{6})$.
\end{center}
Taking the Zariski closure in both ideals, $Z(I_{P_{1}})=\lbrace u_{R} - 1=0 \rbrace   $ and $ Z(I_{P_{2}})=\lbrace u_{A} - 1=0 \rbrace  $, so we have .\\
\begin{center}
$ u_{A}= A*X_{A}=1$ hence we solve for $ X_{A}=A^{-1} $,\\

$ u_{R}=B*X_{R}=1 $, analogously for $ X_{R}=B^{-1} $.
\end{center}
With parameters $ A $ and $ B $ defined below; substituting in the equations with relationship to the repressor $ B*X_{R}=1 $ and activator $ A*X_{A}=1 $ respectively, in the equations,
\begin{center}
$ \dot{X}_{2} =A*X_{A}- (\beta_{2} + \mu)*X_{2}=1-(\beta_{2} + \mu)*X_{2}$,\\
$ \dot{X}_{4} =B*X_{R}- (\beta_{4} + \mu)*X_{4}=1-(\beta_{4} + \mu)*X_{4}$.
\end{center}

And this is a system of first order differential equations with solutions well known, in the literature, and given by:
\begin{center}
% $  X_{2}(t)=(X_{20}+ p(t))*\exp (-(\beta_{2} + \mu)*t)$,\\
% $  X_{4}(t)=(X_{40}+ q(t))*\exp (-(\beta_{4} + \mu)*t)$.
$  X_{2}(t)=X_{20}*\exp (-(\beta_{2} + \mu)*t) +(\beta_{2} + \mu)^{-1} \left(1 -C_{0} \exp (-(\beta_{2} + \mu)*t) \right) $,\\

$  X_{4}(t)=X_{40}*\exp (-(\beta_{4} + \mu)*t) +(\beta_{4} + \mu)^{-1} \left(1 -\tilde{C}_{0} \exp (-(\beta_{4} + \mu)*t) \right)  $.
\end{center} 
With real constants $ X_{20} $ and $ X_{40} $ which are the initial conditions of concentration of mature regulators $ X_{2} $ and $ X_{4} $ respectively, for the solution of the homogeneous differential equation.\\
Also taking the Zariski closure for both ideals $ I_{\sigma^{\vee}_{2}} $ and $I_{\sigma^{\vee}_{4}} $ we have.
\begin{center}
 $Z((I_{\sigma^{\vee}_{2}}))=\left\lbrace  u_{2}u_{A}-1=0, u_{1}-1=0, u_{R}-1=0, u_{3}- 1=0, u_{4}- 1=0   \right\rbrace $.\\
$Z((I_{\sigma^{\vee}_{4}}))=\left\lbrace  u_{4}u_{R}-1=0, u_{1}-1=0, u_{2}-1=0, u_{3}- 1=0, u_{A}- 1=0   \right\rbrace $.

\end{center}
To reproduce the oscillatory phenotype of Savageau, now we study the non-trivial generators, in both Zariski closures above, where to study these kind of solutions associated to Zariski closure, we are understanding, that is a equivalent way to study the steady-state condition associated to each pair $ (Z((I_{\sigma^{\vee}_{i}})),\dot{u}_{i} ) $ for all $ i=1,2,3,4,A,R $; i.e, in our case we want to reproduce the oscillatory phenotype related to the regulators $ u_{2}=(\beta_{2} + \mu)X_{2} $ and $ u_{4}= (\beta_{4} + \mu)X_{4} $ in normalised coordinates. \\
Taking the non-trivial generators above, we have $u_{2}u_{A}-1=0  $ and $  u_{4}u_{R}-1=0$, we can appreciate the inverse relationship between immature and mature regulators, $ u_{A}=u^{-1}_{2} $ and $ u_{R}=u^{-1}_{4}  $, which tell us, most production of mature regulators, less concentration of immature regulators. And substituting in steady-state from the analytical solution given above for $ X_{2}(0) $ and $ X_{4}(0) $. we have $ u_{A}=(\beta_{2} + \mu)^{-1}X^{-1}_{20} $ and $ u_{A}=(\beta_{4} + \mu)^{-1}X^{-1}_{40} $, we substitute in the expression for steady-state equations, $ \dot{X}_{2}=\dot{X}_{4}=0 $, therefore,\\

% solution of the  first differential equation for $ \dot{u}_{1} =u_{2}u_{4}^{-1}- u_{1}$ where the solutions $ u_{2} $ and $ u_{4} $ are the same solutions that of $ X_{2}(t) $ and $  X_{4}(t)$ multiplied only for some constants, once solving this variable, we substitute in the differential equation $\dot{u}_{A} =u_{1}- u_{A}  $, solving for the variable $ X_{A} $ which its solution has a complex value in steady-state, in the same form the differential equation $\dot{u}_{3} =u_{2}- u_{3} $ is solved from solution $ X_{2}(t) $ obtaining the solution for $ u_{3} $, and  so $\dot{u}_{R} =u_{3}- u_{R} $ is solved for $ u_{R} $, multiplying for the constants $ C'_{i} $ we have  $ X_{R} $.\\
 
%from $ u_{3}=1 $ we have a complex solution for the differential equation $ \dot{u}_{R}=1 - u_{R} $, and the same manner from the analytical solution of $ X_{1} $, substituting in the other differential equations, $ \dot{u}_{A}=u_{1}- u_{A} $  obtaining a complex solution too. 

\begin{center}
$ 0=\dot{X}_{2} = (\beta^{0}_{2} + \mu^{0}) \left( \dfrac{\beta_{A}}{\beta^{0}_{A}}\dfrac{\alpha_{1}^{0} n_{2}^{0}}{(\beta_{1}^{0} + \mu^{0})\sqrt{\rho_{1}^{0}} K_{2}^{0}} \right) *(\beta_{2} + \mu)^{-1}X^{-1}_{20} - (\beta_{2} + \mu)*X_{20} $,\\

$ 0=\dot{X}_{4}=(\beta^{0}_{4} + \mu^{0}) \left( \dfrac{\beta_{R}}{\beta^{0}_{R}}\dfrac{\alpha_{3}^{0} n_{4}^{0}}{(\beta_{3}^{0} + \mu^{0})\sqrt{\rho_{3}^{0}} K_{4}^{0}} \right) *(\beta_{4} + \mu)^{-1}X^{-1}_{40} - (\beta_{4} + \mu)X_{40} $.

\end{center}
%From this expression, we clear the variables for the concentrations functions $ X_{2} $, $ X_{4} $, we substitute in $ f_{1} $; and we assume solutions with positive real-part in the polar form,  
\begin{center}
%$ X_{R}=Z_{1}=\omega_{0}\exp(\pm i\omega t)= \,\omega_{0} \left[ \cos(\omega t) \pm \imath \sin(\omega t) \right]  \in \, \mathbb{C}^{*}, \, \omega_{0}\, \in \mathbb{R}_{+}$ and\\
%$ X_{A}=Z_{2}=\theta_{0}  \left[ \cos(\theta t) \pm \imath \sin(\theta t) \right] \, \in \, \mathbb{C}^{*} , \, \theta_{0} \, \in \mathbb{R}_{+}$, 

\end{center}
This type of solutions is suggested by M. Savageau \cite{LomnitzSavageau2014}. Now, we have the desired relation between the regulatory capacities: $ (\beta_{2} + \mu) $ and $( \beta_{4}+ \mu) $. Substituting and clearing terms:
\begin{center}
$ (\beta_{4} + \mu)^{2}*X^{2}_{40}(\beta_{3}^{0} + \mu^{0})\beta^{0}_{R}\sqrt{\rho_{3}^{0}} K_{4}^{0} K_{4A}K_{2}^{0}(\beta^{0}_{2} + \mu^{0})\beta_{A}\alpha_{1}^{0} n_{2}^{0}*Z_{1}= (\beta_{2} + \mu)^{2}*X^{2}_{20}*(\beta_{1}^{0} + \mu^{0}) \sqrt{\rho_{1}^{0}} K_{2}^{0} K_{2} K_{4}^{0} \beta_{R} \alpha_{3}^{0} n_{4}^{0} (\beta^{0}_{4} + \mu^{0}) *Z_{2}$.
\end{center}
clearing $( \beta_{4}+ \mu) $ in the left side and $ (\beta_{2} + \mu) $ in the right side, and after applying logarithms in both sides, we have the expression:

\begin{center}
$ \log_{10} \left( \beta_{4}/ \mu+ 1 \right) =\log_{10} \left\lbrace  \dfrac{X_{40}}{X_{20}} \sqrt{\tilde{K}} \left(  \beta_{2}/ \mu + 1 \right) \right\rbrace  = \log_{10}( \beta_{2}/ \mu + 1 ) + \log_{10} \left(  \bar{Z} \sqrt{\tilde{K}}\right)  $.
\end{center}
This geometrical region can be seen in Figures 3-9; \cite{LomnitzSavageau2014}, given that there is free selection of some parameters (ref. \cite{LomnitzSavageau2014}), the regions change with respect to them; $ \mu, \, \beta_{A}, \, \beta^{0}_{A}, \, \beta_{R}, \, \beta^{0}_{R} $ are free parameters without experimental measure  and the complex number $ \bar{Z}=\dfrac{X_{40}}{X_{20}}=R\exp(\imath*\Theta) \,\in \, \mathbb{C}^{*}$ is given for periodical initial conditions of production regulators, M. Savageau \cite{LomnitzSavageau2014}, here $\tilde{K} $ is defined by:\\\\
$ \tilde{K}=\left( \dfrac{B}{A} \right) \left( \dfrac{K_{2R}}{K^{0}_{2}} \right) \left( \dfrac{K^{0}_{4}}{K_{4A}} \right)  \left(  \dfrac{ \beta^{0}_{4} + \mu^{0} }{ \beta^{0}_{2} + \mu^{0} } \right)  $, where;\\\\

$ A= \left[ \left( \dfrac{\beta_{A}}{\beta^{0}_{A}} \right)    \dfrac{\alpha_{1}^{0} n_{2}^{0}}{(\beta_{1}^{0} + \mu^{0})\sqrt{\rho_{1}^{0}}K_{2}^{0}} \right] $,\\\\

$ B=  \left[ \left( \dfrac{\beta_{R}}{\beta^{0}_{R}} \right) \dfrac{\alpha_{3}^{0} n_{4}^{0}}{(\beta_{3}^{0} + \mu^{0})\sqrt{\rho_{3}^{0}} K_{4}^{0}} \right]$.\\\\

Substituting some parameters for empirical values:, for details, see reference, J. Lomintz and M. Savageau \cite{LomnitzSavageau2014}. We have:\\\\

$ A= \left[ \left( \dfrac{\beta_{A}}{\beta^{0}_{A}} \right)    \dfrac{33.3 mRNAs (30)}{\sqrt{100}(1.5811)} \right] $,\\\\

$B=  \left[ \left( \dfrac{\beta_{R}}{\beta^{0}_{R}} \right) \dfrac{33.3 mRNAs(30)}{\sqrt{100} (0.5)} \right]$.\\\\
$ \dfrac{\beta_{A}}{\beta^{0}_{A}}=0.25 $ and $  \dfrac{\beta_{R}}{\beta^{0}_{R}}=0.175 $ (free values), see \cite{LomnitzSavageau2015}, \cite{LomnitzSavageau2014}, $ K^{0}_{2} \approx \sqrt{10} K^{0}_{4} $, $ K^{0}_{4} \ll 17.3938 $. Every pair of values $ K_{2R} $ and $ K_{4A} $ are calculated with respect to $ \pi_{i}=1, 0 $, $ \delta_{i}=1,0 $ and the formulas given at the begin of the example; also these values and formulas have a bit different normalization, but the intention here is to show that using common toric ideals in the S-system, it is possible to find an invariant region in this case the use of toric ideals, as was the case for $ \langle {u}_{A} - 1 \rangle  $ and $ \langle {u}_{R} - 1 \rangle $.\\\\

In the figures presented (Figures 3 -9), the rigth panel shows yellow geometric regions that reproduce the sustained oscilatory region found for different control modes, which are given by the condition lattice vector $ (\delta_{1},\delta_{3},\pi_{1},\pi_{3}) \,\in \, \mathbb{Z}^{2} $; shown in, J. Lomintz and M. Savageau \cite{LomnitzSavageau2014}. \\
Now we consider other analytical solutions for other phenotypic polytopes, as is the case associated to the phenotypic toric variety $ X_{\sum} =\left\lbrace u_{2}=u^{2}_{4} \right\rbrace$ where this affine space associated to the steady-state solution in all the S-System, and is possible to calculate new analytical expression from this expression, as is shown to continuation, we consider the condition of steady-state is always satisfied:
\begin{center}
$ X_{2}(0)=\tilde{K}*(\beta_{2}+ \mu)*(\beta_{4}+ \mu)X_{4}^{2}(0)  $,
\end{center}
And finally we obtain in logarithmic-space an analytical function defined in a branch of the complex logarithm, i.e., $ \log_{10}(\beta_{4}/\mu) :\mathbb{R}^{+} \longrightarrow \mathbb{C}$. where,
\begin{center}
$ \log_{10}(\beta_{4} / \mu + 1)= \log_{10} \left\lbrace \dfrac{X_{20}}{\tilde{K}*\mu^{2}*X_{4}^{2}(0)}\right\rbrace  - \log_{10}(\beta_{4} / \mu + 1)  $.
\end{center}
In its polar form, it is the case when the value of $ \tilde{\theta} $ has complex values, and in this way we have associated the sustained oscillatory complex region:
\begin{center}
$ \log_{10}(\beta_{4}/\mu + 1)= \log_{10}\left[\tilde{R} \left( \cos(\tilde{\theta}) \pm \imath \sin(\tilde{\theta})  \right) \right]  $.
\end{center}
Where the parameter $ \tilde{K} $ depend of all control modes of oscillatory gene circuit studied, $ \pi^{i}=\delta_{i}=1,0 $ with combinations given in the figures 3-9 for $i=1,2,3,4,A,R $ and  $ \tilde{K} $ is varying in these numbers, each region is visualised in the respective figures. Also the parameters $ \tilde{R} $ and $\tilde{\theta} $ are functions of the parametric variable $(\beta_{2}/\mu + 1)$.

%where $ u_{2}=(\beta_{2}+ \mu)X_{2} $ and $u_{4}= (\beta_{4}+ \mu)X_{4}  $, it follows:
Here we consider constants initial conditions $X_{2}(0)  $ and  $ X_{4}^{2}(0) $ from solution of differential equations, and also to recover the sustained oscillatory region obtained for Savageau we give nominal values for its graphical representation.\\

This expression to point us what an oscillatory sustained region be preserved for, a possible case of biological interpretation, could be that the initial concentration $ C_{0} $ of mature regulator $ X_{2} $ must be constant, as is the case of $  u_{A}=1 $ where was associated the non homogeneous solution of the ordinary differential equation to $ \dot{X}_{2} $, with initial condition $C_{0}$, and the initial concentration of immature regular $ X_{20} $, must be null or almost zero, and $1 \approx \, C_{0}$ must be true.\\

A lattice diagram of toric ideals as subgroups can be done; where the generators $ <u_{A}-1> $ and $ <u_{R}- 1> $ associated to the oscillatory phenotype found in Savageau can be localized in such diagram, respect to a partial order, for exmaplle inclusion of sets. The rest of generators are not yet explored and they also were found  with Hilbert basis as realization of a toric variety $ X_{\sum} $, and will be explored in detail in a future analysis.

\section{Comments and future developments.}
%By now, with these mathematical tools exposed here, we are enabling for the study of S-Systems in steady state by means of computing of toric ideals, since minimal generators contains the solution space of S-Systems, in this way the use of Hilbert basis is important for the solution of these systems, even more with the Theorem \ref{Theorem 2.} give us a clear perspective for the study of dynamical solutions, blowing-up singularities in the parameter-space on genetic-metabolic networks and at the hand with Hilbert basis associated to the lattice polytope in S-Systems, in a next work, we model large-scale genetic-metabolic networks.\\
%Also we will study in a future with more details the concepts, of phenotypic toric variety and phenotypic toric ideals, since they which are associated to sustained oscillating features in the periods of time, under environmental stress.
The mathematical results here presented will enable  the study of S-Systems in steady state by means of computing toric ideals, since minimal generators contain the solution space of S-Systems. In this way the use of Hilbert basis is important for the solution of these systems, even more with Theorem 4.5 which gives a clear perspective for the study of dynamical solutions on genetic-metabolic networks at  hand with Hilbert basis associated to the lattice polytope in S-Systems. These will enable us in a future work,  to model large genetic- metabolic networks. 
%\ref{Theorem 2.} 
\section{Acknowledgment}
We acknowledge funding from UNAM and from the National Institutes of Health (grant number 5R01GM110597-03). Also we give the thanks for her valuable comments, and appreciate all the help and assistance proportioned by the Dr. Margareta Boege for the improvement of this work.

\section{Conclusions.} 
%The principal conclusion of this work is the use of \textbf{Theorem \ref{Theorem 2.3.0.} }; and the \textbf{Hironaka-Atiyah's Theorem} for the demonstrations given in, Theorem \ref{Theorem 1.}-(\textbf{embedding of S-Systems in toric varieties}), and Theorem \ref{Theorem 2.}-(\textbf{toric resolution on S-Systems}), that are fundamentals for the formalization \textbf{genetic-metabolic networks} in the approach studied by, M.A. Savageau, \cite{MASavageau1976}, \cite{MASavageauVoit1987}, \cite{MASavageauVoitIrvine987}, by means of mathematical modelling in S-Systems. These facts show that the S-Systems are toric varieties. And as consequence of these results, the Theorem \ref{Theorem 1.} give us a formalization for the study of S-Systems in steady-state. But of a more general way, the Theorem \ref{Theorem 2.} opens for the \textbf{dynamical study} of genetic-metabolic networks, in both we have seen the use of \textit{Hilbert basis} for the realizations of toric morphisms, but computationally it opens to us the possibility for the mathematical modelling on \textit{large-scale genetic-metabolic networks}; since the algorithms for computing of Hilbert basis in \textit{high-dimensional lattice polytope} are possible.\\ 

The main result in this work is the use of \textbf{Theorem 3.6}, and the \textbf{Hironaka-Atiyah's Theorem} for the demonstrations given in Theorems, Theorem 4.4-(\textbf{embedding of S-Systems in toric varieties}), and Theorem 4.5-(\textbf{toric resolution on S-Systems}), that give the foundation to the toric formalization of genetic-metabolic networks in the approach studied by M.A. Savageau. These facts demonstrate that S-Systems are toric varieties. As a consequence of these results, the Theorem 4.4 give us a formalization for the study of S-Systems in steady-state. In more general terms, the Theorem 4.5 provides the formal support to the use of toric algebraic geometry tools to the study of the dynamics of genetic-metabolic networks. In both cases we have seen the use of \textit{Hilbert basis} for the realizations of toric morphisms, and computationally it opens the possibility for this mathematical modelling of large-scale genetic-metabolic networks, since the algorithms to compute Hilbert basis in high-dimensional lattice polytope can be used. Also we analyse a particular genetic-metabolic network, computing the toric ideals and toric varieties associated to the S-system of this network, this network is the case of a synthetic oscillating design, studied by Savageau \cite{LomnitzSavageau2014} in the which was reported the existence of a sustained oscillating region (oscillating phenotype) when the regulation  transcriptional was submitted under different control modes, we reproduced this region with an ideal toric common called phenotypic toric ideal, and we also constructed the phenotypic toric variety, where we can conclude that these geometrical-algebraic objects are invariant too under these control modes, finding the possibility of more intrinsic properties in the topology of any gene-metabolic network, and to perform future studies of these structures.


\begin{thebibliography}{1}


%%%%\bibitem{AtiyahMacDonald1969}
%%%%M.~Atiyah and I. G.~MacDonald.
%\newblock Information theory and an extension of the maximum likelihood
 % principle.
%%%%{\em Introduction to Conmmutative Algebra.}, Addison-Wesley, Reading, MA, 1969.




%%%\bibitem{MAoyaguiWatanabe2005}
%%%M.~Aoyagui and S.~Watanabe.
%%%\newblock Resolution Of Singularities And Generalization Error With Bayesian Estimation For Layered Neural Network.
%%%{\em Neural Networks.} Vol.J88-D-II, No.10, pp.2112-2124, 2005.
%%%\bibitem{MAoyaguiWatanabe2005}
%%%%M.~Aoyagui and S.~Watanabe.
%%%%\newblock Generalization  Error Of Three Layered Learning Model In Bayesian Estimation.
%%%%{\em Proceedings Of The IASTED International Conference Computational Intelligence} San Francisco USA, pp.20-22, Nov. 2006.
%%%%%%%\bibitem{MFAtiyah1970}
%%M.F.~Atiyah.
%\newblock Information theory and an extension of the maximum likelihood
 % principle.
%%{\em Resolution of Singularities and Division of Distributions.} Communications on Pure and Applied Mathematics, Vol. XXIII, 145-150, 1970.


%%%%%\bibitem{ABronsted1983}
%%%A.~Bronsted.
%\newblock Information theory and an extension of the maximum likelihood
 % principle.
%%%%{\em An Introduction to Convex Polytopes.} Springer-Verlag, New York 10010, 1983.
%%\bibitem{CMBishop1987}
%%%C. M.~Bishop.
%\newblock Information theory and an extension of the maximum likelihood
 % principle.
%%%{\em Neural Networks And Machine Learning.} Serie F: Computer and Systems Sciences, Springer, Vol. 168, Cambridge CB2 3NH, U.K, 1997.
\bibitem{MFAtiyah1970}
M.F.~Atiyah.
%\newblock Information theory and an extension of the maximum likelihood
 % principle.
{\em Resolution of Singularities and Division of Distributions}, Communications On Pure and Applied Mathematics, vol. XXIII, 145-150 (1970).

\bibitem{MPCastilloOscar2017}
M.P.~Castillo-Villalba and J.O.~Gonzalez-Cervantes.
%\newblock Information theory and an extension of the maximum likelihood
 % principle.
{\em An Approach with Toric Varieties for Singular Learning Machines}, arXiv:1708.02273v1, August, 2017.


%\bibitem{MPCastilloLauraJulio2017}
%Marco Polo~Castillo-Villalba, Laura~Gomez-Romero and Julio~Collado-Vides.
%\newblock Information theory and an extension of the maximum likelihood
 % principle.
%{\em Genetic-metabolic networks can be modeled as toric varieties},  arXiv:1708.05384v1, August, 2017.




\bibitem{DCoxLittleOshea2004}
D.~Cox, J.~Little and D.~O'Shea.
%\newblock Information theory and an extension of the maximum likelihood
 % principle.
{\em Using Algebraic Geometry.} Springer, 2nd. Edition, August, 2004.

%%%%\bibitem{DCoxLittleOshea2004}
%%%%%D.~Cox, J.~Little and D.~O'Shea.
%\newblock Information theory and an extension of the maximum likelihood
 % principle.
%%%{\em Ideals, Varieties, and Algorithms (An Introduction to Computational Geometry and Conmmutative Algebra)} Springer, 2nd. Edition, August, 2004.

\bibitem{DCoxLittleSchenck2010}
D.~Cox, J.~Little and H.~Schenck.
%\newblock Information theory and an extension of the maximum likelihood
 % principle.
{\em Toric Varieties} Department Of Mathematics, Amherst College, Amherst MA 01002, 2010.

\bibitem[DGPS]{DGPS}
Decker, W.; Greuel, G.-M.; Pfister, G.; Sch{\"o}nemann, H.: 
\newblock {\sc Singular} {4-1-0} --- {A} computer algebra system for polynomial computations.
\newblock {http://www.singular.uni-kl.de} (2016).


%\bibitem{LDGarciaStillmanSturmfels2005}
%L. D.~Garcia, M.~Stillman and B.~Sturmfels.
%\newblock Algebraic geometry of Bayesian networks
%{\em Journal of Symbolic Computation.} Volume 39, Issues 3–4, Pages 331-355, March–April 2005.




\bibitem{GEwald1993}
G.~Ewald.
%\newblock Combinatorial Convexity and Algebraic Geometry.
%\newblock In T.M. Fomby and R.~Carter Hill, editors, 
{\em Combinatorial Convexity and Algebraic Geometry.}, Springer-Verlag, New York, Inc., 1996.




%%%%\bibitem{TFukui2000}
%%%%T.~Fukui.
%\newblock Combinatorial Convexity and Algebraic Geometry.
%\newblock In T.M. Fomby and R.~Carter Hill, editors, 
%%%%{\em Introduction to Toric Modifications With an Application to Real Singularities.} Department Of Mathematics, Faculty Of Science, Saitama University, 255 Shimo-Okubo, Urawa 338-8570, Japan, pag. 96-114, 2000.
%%%%\bibitem{MHenkWeismantel1964}
%%%%M.~Henk, R.~Weismantel.
%\newblock Resolution Of Singularities Of An Algebraic Variety Over A Field Of Zero Characteristic.
%\newblock In T.M. Fomby and R.~Carter Hill, editors, 
%%%%{\em On Hilbert Bases Of Polyhedral Cones.} Mathematical Foundations of Computer Science, 1999.




\bibitem{HHironaka1964}
H.~Hironaka.
\newblock Resolution Of Singularities Of An Algebraic Variety Over A Field Of Zero Characteristic.
%\newblock In T.M. Fomby and R.~Carter Hill, editors, 
{\em Annals Of Mathematics.} Vol.79, pp.109-326, 1964.

%%\bibitem{ASPoznyak2009}
%%%A.S.~Poznyak.

%%{\em Advanced Mathematical Tools for Automatic Control Engineers; Vol. 2, Stochastic Techniques.} Elsevier, %%Oxford, UK, 2009.
%%\bibitem{ASPoznyakSanchezYu2001}
%%A.S.~Poznyak, E.N.~Sanchez and W.~Yu.
%\newblock Combinatorial Convexity and Algebraic Geometry.
%\newblock In T.M. Fomby and R.~Carter Hill, editors, 
%%{\em Differential Neural Networks For Robust Nonlinear Control: Identification, State Estimation and Trajectory Tracking} World Scientific, Farrer Road, Singapore 912805, 2001.

\bibitem{DLedezma2017}
Daniela~Ledezma-Tejeida, Cecilia~Ishida and Julio~Collado-Vides.
\newblock Genome-Wide Mapping of Transcriptional Regulation and Metabolism DescribesInformation-Processing Units in
Escherichia coli.
%\newblock In T.M. Fomby and R.~Carter Hill, editors, 
{\em frontiers in Miocrobiology.} Vol. 8, article 1466, 03 August, 2017.






\bibitem{LomnitzSavageau2013}
J. G.~Lomnitz, M. A.~Savageau.
\newblock Phenotypic deconstruction of gene circuitry
%\newblock In T.M. Fomby and R.~Carter Hill, editors, 
{\em Chaos.} Vol. 23, Issue 2, May 21, 2013.
\bibitem{LomnitzSavageau2014}
J. G.~Lomnitz, M. A.~Savageau.
\newblock Strategy Revealing Phenotypic Differences among Synthetic
Oscillator Designs.
%\newblock In T.M. Fomby and R.~Carter Hill, editors, 
{\em ACS Synth. Biol.} Volume 3, 686−701, July 14, 2014.

\bibitem{LomnitzSavageau2015}
J. G.~Lomnitz, M. A.~Savageau.
\newblock Elucidating the genotype-phenotype map by automatic
enumeration and analysis of the phenotypic repertoire.
%\newblock In T.M. Fomby and R.~Carter Hill, editors, 
{\em npj Systems Biology and Applications.} Vol 1, 15003; doi:10.1038/npjsba.2015.3; published online 28 September 2015.
\bibitem{LomnitzSavageau2016}
J. G.~Lomnitz, M. A.~Savageau.
\newblock Design Space Toolbox V2: Automated
Software Enabling a Novel
Phenotype-Centric Modeling
Strategy for Natural and Synthetic
Biological Systems.
%\newblock In T.M. Fomby and R.~Carter Hill, editors, 
{\em Frontiers in Genetics.} Vol 27, article 117, July 12, 2015.

\bibitem{MASavageau1976}
M.A.~Savageau.
\newblock Biochemical Systems Analysis: “A Study of Function and Design in Molecular Biology”.
%\newblock In T.M. Fomby and R.~Carter Hill, editors, 
{\em Biochemical Systems Analysis.}, volume 1. Addison-Wesley Inc , Reading, Mass, USA, 1976.
\bibitem{MASavageauVoit1987}
M. A.~Savageau and E. O.~Voit.
\newblock Recasting Nonlinear Differential Equations as S-Systems:
A Canonical Nonlinear Form.
%\newblock In T.M. Fomby and R.~Carter Hill, editors, 
{\em Mathematical Biosciences.}, Volume 87, Issue 1, pages 83-115, November 1987.
\bibitem{MASavageauVoitIrvine987}
M. A.~Savageau, H. O.~Voit and D. H.~Irvine.
\newblock Biochemical Systems Theory and Metabolic Control Theory:
1. Fundamental Similarities and Differences.
%\newblock In T.M. Fomby and R.~Carter Hill, editors, 
{\em Mathematical Biosciences.}, Volume 86, Issue(2), pages 127-145, October 1987. 
\bibitem{MASavageau2001}
M. A.~Savageau.
\newblock Design principles for elementary gene circuits: Elements, methods, and examples.
%\newblock In T.M. Fomby and R.~Carter Hill, editors, 
{\em Chaos.}, Volume 11, Issue 1, December (2001).

\bibitem{SavageauCoelhoFasaniSalvador2009}
M. A.~Savageau, P. M. B. M.~Coelho, R. A.~Fasani  and A.~Salvador.
\newblock Phenotypes	and	tolerances	in	the	design	space
of	biochemical	systems.
%\newblock In T.M. Fomby and R.~Carter Hill, editors, 
{\em PNAS.} Vol. 106, no. 16, 6435– 6440, April 21, 2009.

\bibitem{BSturmfels1996}
B.~Sturmfels.
\newblock Grobner Bases and Convex Polytopes.
%\newblock In T.M. Fomby and R.~Carter Hill, editors, 
{\em American Mathematical Society.} Providence, RI, 1996.



%%\bibitem{SWatanabe12001}
%S.~Watanabe.
%%\newblock Algebraic Geometry Of Learning Machines With Singularities And Their Prior Distributions.

%%%%%{\em Journal Of Japanese Society  For Artificial Intelligence.} Vol. 16, No. 2; Pages 308-315(2001). 
%%%\bibitem{SWatanabe22001}
%%S.~Watanabe.
%%\newblock Algebraic Information Geometry For Learning Machines With Singularities.
%\newblock In T.M. Fomby and R.~Carter Hill, editors, 
%%{\em Neural Networks.} Vol. 14 Issue 8, pages 1049-1060, January 2001. 

%%%\bibitem{SWatanabe32001}
%%S.~Watanabe.
%%\newblock Training And Generalization Error Of Learning Machines With Algebraic Singularities.
%\newblock In T.M. Fomby and R.~Carter Hill, editors, 
%%{\em IEICE Transactions.} Vol. J84-A, No.1, pp.99-108, Jan. 2001.
%%\bibitem{SWatanabe42001}
%%%S.~Watanabe.
%%%%\newblock Algebraic Geometrical Methods For Hierachical Learning Machines.
%\newblock In T.M. Fomby and R.~Carter Hill, editors, 
%%%%{\em Neural Networks.} Vol. 14 Issue 8, pages 1049-1060, Oct. 2001.
%%%\bibitem{SWatanabe52001}
%%%S.~Watanabe.
%%%\newblock Algebraic Analysis For Non-Identifiable Learning Machines.
%\newblock In T.M. Fomby and R.~Carter Hill, editors, 
%%%%%{\em Neural Computation.} Vol. 13, No. 4, pp.899-933, 2001.

%%%%\bibitem{SWatanabe62001}
%%%S.~Watanabe.
%%%%\newblock Learning Efficiency Of Redundant Neural Networks In Bayesian Estimation.
%\newblock In T.M. Fomby and R.~Carter Hill, editors, 
%%%%{\em IEEE Transactions On Neural Networks.} Vol.12, No.6, pp.1475-1486, 2001.
%%%%\bibitem{SWatanabe72001}
%%S.~Watanabe.
%%%\newblock Learning Efficiency Of Redundant Neural Networks In Bayesian Estimation.
%\newblock In T.M. Fomby and R.~Carter Hill, editors, 
%%%{\em IEEE Transactions On Neural Networks.} Vol.12, No.6, pp.1475-1486, 2001.

%%%\bibitem{SWatanabeYamazaki2002}
%%%S.~Watanabe and K~Yamazaki.
%%%\newblock A Probabilistic Algorithm to Calculate The Learning Curves Of Hierachical Learning Machines With Singularities.
%\newblock In T.M. Fomby and R.~Carter Hill, editors, 
%%%{\em Trans. On IEICE.} Vol. J85-D-II, No.3, pp.363-372, Mar. 2002.
%%%\bibitem{SWatanabe2005}
%%%S.~Watanabe.
%%%\newblock Algebraic Geometry Of Singular Learning Machines And Symmetry Of Generalization And Training Errors.
%\newblock In T.M. Fomby and R.~Carter Hill, editors, 
%%%{\em Neurocomputing.} Vol. 67, Pages 198-213, August 2005.

%%%\bibitem{SWatanabe2019}
%%%S.~Watanabe.
%%%\newblock Almost All Learning Machines Are Singular.
%\newblock In T.M. Fomby and R.~Carter Hill, editors, 
%%%{\em Invited Paper in FOCI2007.} 2009.
%%\bibitem{SWatanabe2009}
%%S.~Watanabe.
%%\newblock Algebraic Geometry And Statistical Learning Theory.
%\newblock In T.M. Fomby and R.~Carter Hill, editors, 
%%%{\em Cambridge Monographs On Applied And Computational Mathematics.} Cambridge CB2 8RU, U.K, 2009.




%%%\bibitem{KYamazakiWatanabe2003}
%%K.~Yamazaki and S.~Watanabe.
%%%\newblock Singularities In Mixture Models And Upper Bounds Of Stochastic Complexity.
%\newblock In T.M. Fomby and R.~Carter Hill, editors, 
%%{\em Neural Networks.} Vol. 16, Issue 7, pages 1029-1038, September 2003.

%%%%\bibitem{KYamazakiWatanabe2004}
%%%%K.~Yamazaki, M.~Aoyagui and S.~Watanabe.
%%\newblock Stochastic Complexity And Newton Diagram.
%\newblock In T.M. Fomby and R.~Carter Hill, editors, 
%%%{\em International Symposium On Information Theory And Its Applications.} Parma Italy, pp.10-13, Oct. 2004. 

\end{thebibliography}
\end{document}